\documentclass[aos]{imsart}

\RequirePackage{amsthm,amsmath,amsfonts,amssymb,bbm}
\RequirePackage[authoryear]{natbib}%
\RequirePackage{xcolor}
\RequirePackage[colorlinks,citecolor=blue,urlcolor=blue]{hyperref}%
\RequirePackage{graphicx}%
\usepackage{subfigure}
\usepackage{float}
\usepackage{multirow}
\usepackage{diagbox}
\usepackage[figuresright]{rotating}
\usepackage{pdflscape}
\usepackage{booktabs}
\usepackage{bm}
\usepackage{algorithm}
\usepackage{algorithmic}
\usepackage{changepage}
\usepackage{tikz}
\startlocaldefs
\theoremstyle{plain}

\newtheorem{theorem}{Theorem}[section]

\newtheorem{proposition}{Proposition}[section]

\theoremstyle{definition}

\newtheorem{assumption}{Assumption}
\newtheorem{remark}{Remark}[section]

\newcounter{cnstcnt}

\newcounter{bnstcnt}

\newcounter{dbnstcnt}

\newcounter{dcnstcnt}

\makeatletter
\newcommand*{\centerfloat}{%
  \parindent \z@
  \leftskip \z@ \@plus 1fil \@minus \marginparwidth
  \rightskip \leftskip
  \parfillskip \z@skip}
\makeatother

\usepackage{enumerate}
\usepackage{lscape}
\usepackage{threeparttable}  %

\newcommand{\bo}{\beta_{0}}

\newcommand{\ao}{\alpha_{0}}

\newcommand{\var}{\operatorname{Var}}
\newcommand{\nux}{\nu_x}
\newcommand{\nuy}{\nu_y}
\newcommand{\Ltwo}{\mathcal{L}^2}

\newcommand{\ppp}{\mathbb{P}}
\newcommand{\T}{\Gamma}
\newcommand{\Tnm}{ \widehat{\Gamma}}
\newcommand{\Pinm}{ \widehat{\Pi}}
\newcommand{\Pix}{ \Pi_x}
\newcommand{\Piy}{ \Pi_y}

\endlocaldefs

\begin{document}

\begin{frontmatter}
\title{Functional linear regression from sparse to dense designs: a pooling-ridge method and minimax optimality}
\runtitle{Discretely observed FLR}

\begin{aug}
\author[A]{\fnms{Shunxing}~\snm{Yan}\ead[label=e1]{sxyan@tsinghua.edu.cn}}
\author[B]{\fnms{Fang}~\snm{Yao}\ead[label=e2]{fyao@math.pku.edu.cn}}

\address[A]{Department of Statistics and Data Science, Tsinghua University
\printead[presep={ \ }]{e1}}
\address[B]{School of Mathematical Sciences,
Center for Statistical Science, Peking University 
\printead[presep={ \ }]{e2}}
\end{aug}

\begin{abstract}
Functional data analysis is an important statistical field that treats data as random functions. In practice, the random functions are often not fully observed but instead measured at discrete times. While simpler problems, such as mean and covariance estimation, have been widely studied for discretely observed data, optimal estimation of linear regression for this data type has remained unsolved for over two decades. To tackle this fundamental challenge, we propose a novel approach, referred to as pooling ridge estimation, which combines the advantages of pooling strategy and RKHS-based method by incorporating the unbiased estimation of operators based on discretely observed measurements from all subjects. This unified estimation framework enables us to achieve minimax optimality in prediction risk in arbitrary sampling schemes ranging from sparse to dense designs, for both scalar-on-function and function-on-function regression models.
Such methodological and theoretical advances are obtained for the first time and accurately reveal the influence of discrete sampling. For scalar-on-function regression, the phase transition occurs once, separating the convergence behavior into two distinct regimes. Remarkably, for function-on-function regression, up to three phase transitions may occur, determined by the sampling frequencies of the predictor/response functions. Finally, simulation experiments and two real data examples provide empirical support for the proposed methods.
\end{abstract}

\begin{keyword}[class=MSC]
\kwdgroup[type=primary]{\kwd{62R10}}
\kwdgroup[type=secondary]{\kwd{62G08}\kwd{62C20}}
\end{keyword}

\begin{keyword}
\kwd{Functional data}
\kwd{phase transition}
\kwd{pooling}
\kwd{RKHS}
\end{keyword}

\end{frontmatter}

\section{Introduction}
\subsection{Literature review and open problems}
Functional data analysis (FDA) has gained significant attention in the past several decades, reflecting the importance of analyzing data in the form of functions over some domain, such as time or space. 
This methodological framework has been extensively applied in various fields, including systems biology, econometrics, chemoinformatics, climate modeling, environmental statistics, and so on. 
A substantial body of literature has been dedicated to functional data analysis; for instance, see the monographs \citet{ramsay2005,ferraty2006nonparametric,hsing2015theoretical,kokoszka2017introduction} and the survey papers 
\citet{wang2016functional,aneiros2019recent} for a comprehensive overview. 
Despite significant progress and theoretical development in this area, several critical challenges remain unresolved. One of those is the minimax optimal estimation for linear regression with discretely observed functional data, ranging from sparse to dense designs, which is the primary focus of this paper.

Functional linear regression is a widely employed model that characterizes the relationship between a functional predictor and some response variable through a linear operator.
When the response variable is scalar, the model is referred to as scalar-on-function regression, while function-on-function regression refers the case when the response is also functional.
The slope function in these problems is typically estimated using data-driven bases or within specified function spaces.
For scalar-on-function regression, \citet{cai2006prediction} and \citet{hall2007methodology} investigated optimal estimation based on functional principal component analysis (FPCA). 
To address the nonlinearity challenges in exponential family models, \citet{dou2012estimation} proposed a change-of-measure argument and established the minimax optimal result. 
Within the reproducing kernel Hilbert space (RKHS) framework, \citet{yuan2010reproducing} and \citet{cai2012minimax} developed the estimation procedure and derived minimax optimal prediction errors.
Furthermore, many other works, such as \citet{du2014penalized}, \citet{qu2016optimal}, and \citet{balasubramanianfunctional}, extended different model settings with scalar responses. 
Function-on-function regression presents greater challenges due to the complexity of the response and therefore has less literature. \citet{yao2005b} allow the functions to be sparsely observed and \citet{Christophe2013} derived the prediction error for fully observed data. 
Further studies, like \citet{luo2017function}, proposed various data-driven methods. Using the RKHS tools, \citet{lian2015minimax} and \citet{sun2018optimal} developed estimators and established their theoretical properties.
Overall, whether the response is a scalar or a function, functional linear regression has been extensively studied, with both data-driven and space-specific (e.g., RKHS) methods providing well-established estimation strategies.

Despite substantial progress, the existing literature exhibits several limitations from both methodological and theoretical perspectives. 
The assumption of fully observed functional data is often unrealistic and represents an ``idealized'' scenario. In practice, one typically observes \(n\) (referred to as sample size) independently and identically distributed (i.i.d.) random functions, with the \(i\)-th function intermittently measured at \(m_{x,i}\) discrete times (referred to as sampling frequency). 
From a methodological standpoint, space-specific works \citep{cardot2003spline,crambes2009smoothing,james2009functional,yuan2010reproducing,zhang2013time,Zhang02012020,dette2024statistical} rarely account for discrete, especially sparse, observations.
Existing RKHS-based methods, including those for scalar-on-function regression \citep{yuan2010reproducing,cai2012minimax} and function-on-function regression \citep{lian2015minimax,sun2018optimal}, are designed under the assumption of fully observed trajectories. 
To apply these methods in more realistic cases, a common practice is pre-smoothing, which involves reconstructing the curve of each function individually from its discrete observations. 
However, this approach can introduce significant bias when functions are sparsely sampled \citep{yao2005a}. In such settings, better data efficiency can be achieved by pooling observations across subjects to estimate marginal quantities. For example, several FPCA-based methods \citep{yao2005b, zhou2023functional} employ pooling strategies specifically designed for discretely observed functional data in linear models.
There have also been some related attempts to study regression with sparsely observed functional data using RKHS tools. For instance, \citet{mostafaiy2019optimal} proposed a function-on-function linear regression estimator for sparse noisy data, based on RKHS-regularized estimation of covariance and FPCA, followed by an FPCA-based regression step. However, the corresponding theoretical analysis with a diverging number of components is unavailable. 
In addition, \citet{sang2022nonlinear} studied nonlinear function-on-function regression. For sparse observations, their approach first recovers individual trajectories under an FPCA framework and then applies the regression method to the recovered functions; the impact of the sampling frequency on the estimates is still unknown.
Nevertheless, direct RKHS-based methods and minimax theory for discretely observed functional linear regression remain much limited.

From the theoretical perspective, existing literature on minimax optimal convergence for functional linear models \citep{cai2006prediction,hall2007methodology,dou2012estimation,yuan2010reproducing,cai2012minimax,lian2015minimax,sun2018optimal} has focused exclusively on fully observed functional data, but have not accounted for the impact of sampling frequency. 
In contrast, many works have investigated the influence of discrete observations on the mean and covariance estimation 
\citep{Cai2010NonparametricCF,cai2011optimal,li2010uniform, zhang2016sparse,wang2022low,yan2025drrm}.
They reveal an interesting phase transition phenomenon in the convergence rate which occurs when the sampling frequency reaches a certain order relative to the sample size. 
Optimal estimation for linear regression of discretely observed functional data remains unknown since \citet{yao2005b,yao2005a} due to the difficulty of an infinite-dimensional inverse problem complicated by discrete and noisy measurements. 
Similar phase transition phenomena have been widely conjectured, but there have been no significant theoretical advancements over the past decades. The theoretical challenge arises since the techniques and results for fully observed data \citep{hall2007methodology,yuan2010reproducing} cannot be directly adopted, especially in sparse sampling design.
As far as we know, the most related work is \citet{zhou2023functional}. They investigate the scalar-on-function regression discretely observed data and establish the convergence rate under sufficiently dense sampling. 
However, since the phase transition boundary has not been precisely characterized, their requirement of dense design imposes unnecessarily high-order sampling frequency. 
Recent work \citet{zhou2024theoryfunctionalprincipalcomponent} improves the theoretical results on FPCA, but is not adequate to establish the minimax optimal rate of linear models with various sampling frequencies. 
Furthermore, no work has obtained complete results for the function-on-function regression problem, which should simultaneously address the challenges from the sampling frequencies of both predictor and response functions. 
Overall, these facts confirm the substantial  challenges in establishing optimal estimation and theory for linear models, with discretely observed functional data in both sparse and dense sampling designs.

\subsection{Motivations and our contributions}\label{sec:motcon}
In this study, for the functional linear regression, we aim to bridge the gap between the idealized scenario of fully observed functions and reality with only discrete and noisy observations, by proposing an RKHS-based pooling method. 
The new estimator should unify arbitrary realistic sampling schemes ranging from sparse to dense designs, as well as accommodate both scalar-on-function and function-on-function regression models.
For simplicity, we assume that the sampling frequencies of all predictor functions are in the same order as \(m_x\). In function-on-function regression, the sampling frequencies of all response functions are assumed to be of order \(m_y\).
As stated before, existing methods based on RKHS \citep{yuan2010reproducing,sun2018optimal} are mainly developed for fully observed functional data, and pre-smoothing does not work in sparse design. 
A possible remedy is to approximate the required integrals in the loss functions of such methods for fully observed data, using discrete observations, but this also introduces a bias of \( O(m_x^{-1}) \), which becomes substantial for small sampling frequency \( m_x \).
Alternatively, FPCA-based estimation \citep{yao2005b, zhou2023functional} pools observations from all subjects to estimate the covariance function and, consequently, the desired principal component bases. While the theoretical analysis of principal component estimation is still challenging, the pooling strategy improves the estimation efficiency for sparsely observed samples.
Building on these insights and to accommodate various sampling frequencies, we combine the advantages of the RKHS methods and the pooling strategy in FPCA estimation. 
Instead of estimating covariance function, we pool discretely observed measurements from all subjects to obtain unbiased estimators for operators in RKHS methods and plug them to obtain final estimation. 
This is a unified framework applicable to both scalar-on-function and function-on-function regression models, referred to as the pooling-ridge estimator.
We then establish the convergence rates for the proposed estimators, even in misspecified cases, i.e. allowing the slope function to lie outside the considered RKHS or exhibit higher regularity with faster coefficient decay. Additionally, matching lower bounds are derived, thereby demonstrating the optimality of our estimators in terms of norm induced by prediction.

To intuitively illustrate the theoretical findings, we introduce toy examples where the predictor functions share aligned eigenfunctions with the given kernel. For more general and formal conclusions applicable to practical settings, please see Section \ref{sec:theory}.
Throughout, we use the notation \( a_n \lesssim b_n \) (or \( a_n \gtrsim b_n \)) to indicate \( a_n \leq C b_n \) (or \( Ca_n \geq b_n \)) for some constants \( C > 0 \). Additionally, \( a_n \asymp b_n \) denotes that both \( a_n \lesssim b_n \) and \( b_n \lesssim a_n \) hold. 
Suppose that the slope function \( \bo \) resides in the RKHS \( \mathcal{H}(K) \) associated with a kernel \( K \), where the kernel \( K \) is assumed to share aligned eigenfunctions with the covariance function \( C \) of predictor \( X \).
Assume that the simple eigenvalue decay rates of \( C \) and \( K \) are \( k^{-2r_c} \) and \( k^{-2r_b} \), respectively. Intuitively, \(r_c\) and \(r_b\) are the ``smoothness'' of predictor functions and slope functions.
In scalar-on-function regression, the minimax optimal rate in terms of prediction is \[n^{-\frac{2r_b + 2r_c }{2r_b + 2r_c  + 1}} + (nm_x)^{-\frac{2r_b + 2r_c }{2r_b + 4r_c  + 1}},\]
where the second term reflects the impact of discrete observations and is identified for the first time in this work.
Here, a phase transition occurs at order \(m_x \asymp n^{2r_c/(2r_b+2r_c+1)}\). %
Specifically, in the dense design defined as when the sampling frequency \(m_x \gtrsim n^{2r_c/(2r_b+2r_c+1)}\), %
the convergence rate reaches \(n^{-(2r_b + 2r_c)/ (2r_b + 2r_c  + 1)}\). %
Further increases in the sampling frequency do not improve the rate, as \( n^{-(2r_b + 2r_c)/(2r_b + 2r_c + 1)} \) %
represents a fundamental limit, even in the ideal case where all \( n \) functions are fully observed \citep{yuan2010reproducing,dou2012estimation}. On the contrary, for sparse sampling where \( m_x \lesssim n^{2r_c/(2r_b+2r_c+1)}\), %
the convergence slows to a rate of order \( (nm_x)^{-(2r_b + 2r_c)/(2r_b + 4r_c + 1)} \), %
which depends on the total number of observation points \( nm_x \).
An interesting phenomenon, termed ``{\it curse of smoothness}'' occurs at the phase transition boundary. This refers to the fact that increasing \( r_c \) reduces model complexity but delays the phase transition, a distinctive property of functional regression problems from mean and covariance estimations.

\begin{figure}[htbp]
	\centering
	\begin{minipage}{0.49\linewidth}
		\centering
		\includegraphics[width=0.9\linewidth]{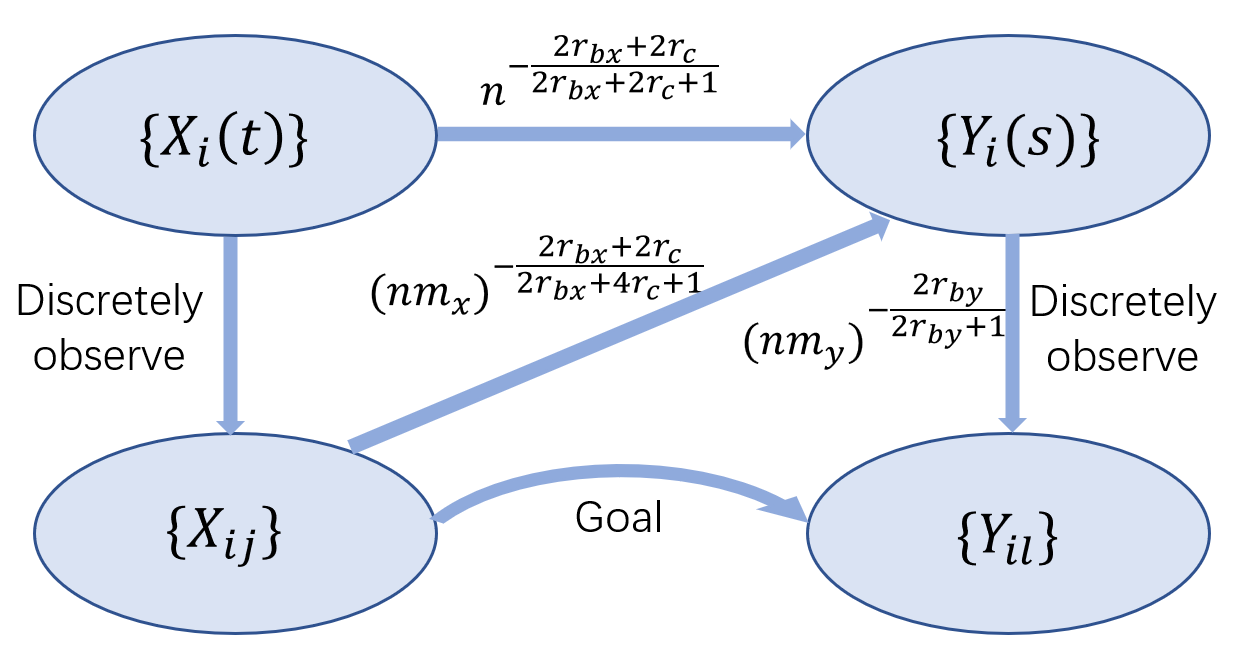}
	\end{minipage}
    \caption{ The intuitive implications of each term in minimax optimal rate of function-on-function regression in terms of norm induced by prediction, under the aligned case. The term \(n^{-(2r_{bx} + 2r_c)/(2r_{bx} + 2r_c  + 1 )}\) is determined by \(n\) pairs functions; while 
    \( (nm_x)^{-(2r_{bx} + 2r_c )/(2r_{bx} + 4r_c + 1)} \) and
    \( (nm_y)^{-2r_{by}/(2r_{by} + 1)}\) are introduced by the discrete measurements of the predictor and response functions, respectively. 
    }\label{fig:shiyi}
\end{figure}

\begin{figure}[htbp]
	\centering
	\begin{minipage}{0.34\linewidth}
		\centering
		\includegraphics[width=0.95\linewidth]{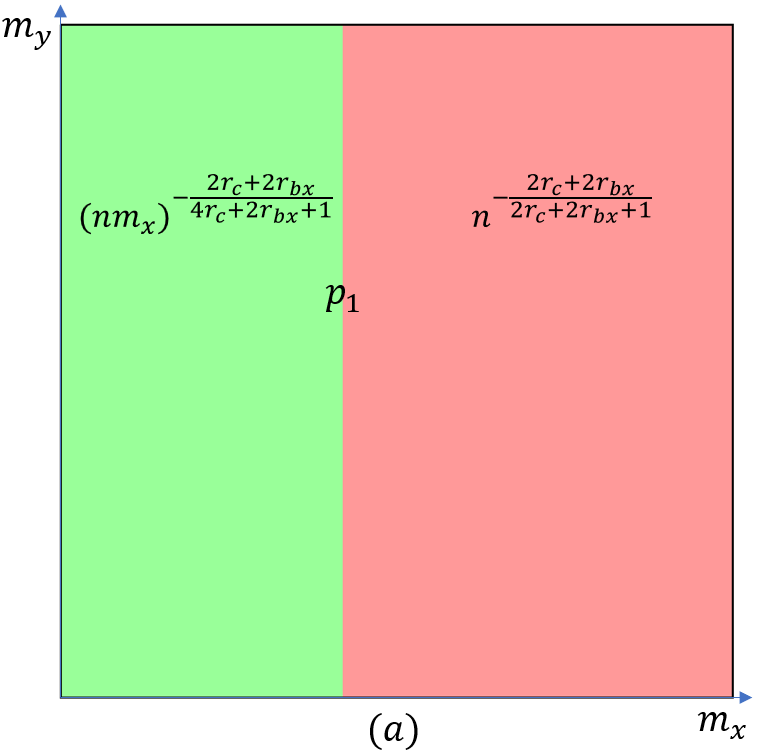}
	\end{minipage}
	\begin{minipage}{0.34\linewidth}
		\centering
		\includegraphics[width=0.95\linewidth]{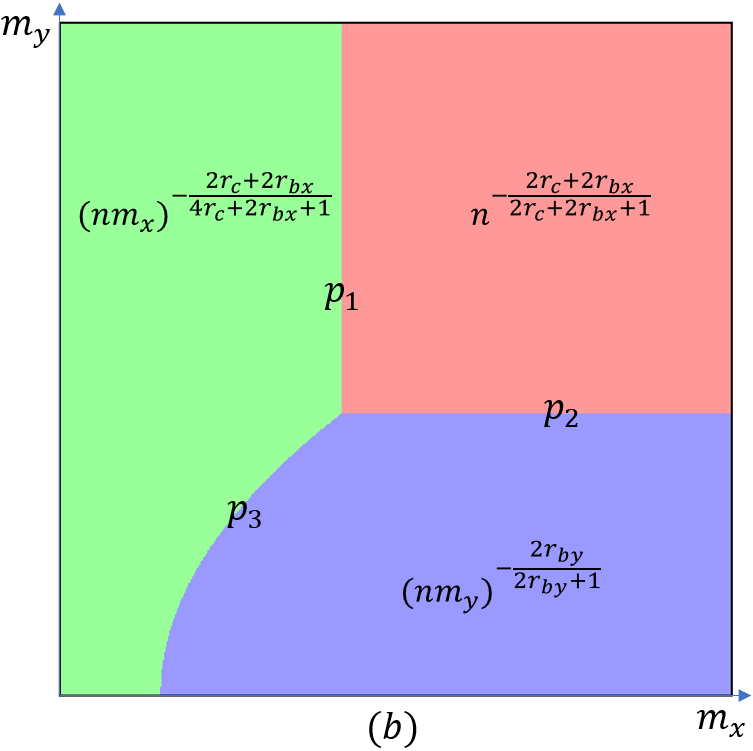}
	\end{minipage}
    	\begin{minipage}{0.34\linewidth}
		\centering
		\includegraphics[width=0.95\linewidth]{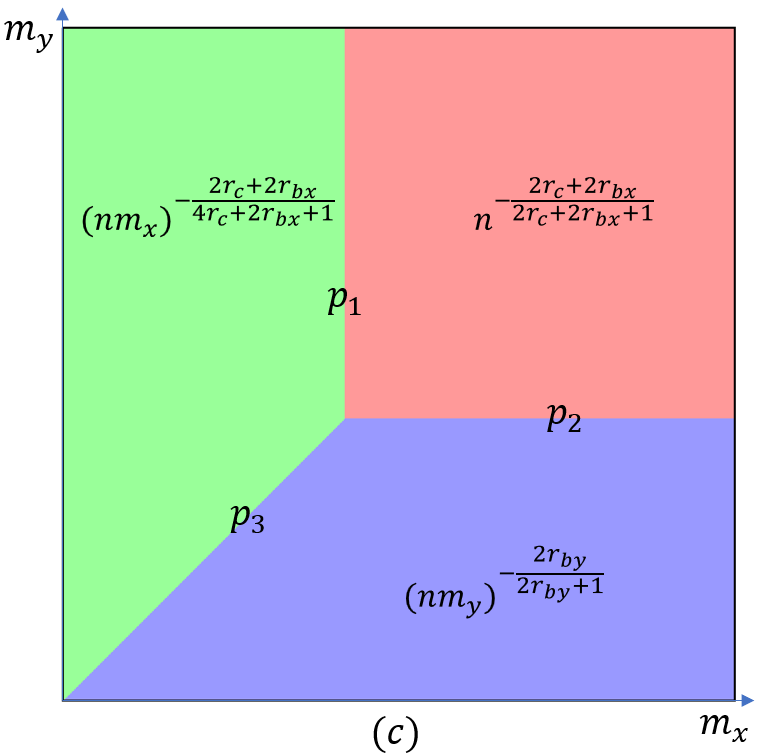}
	\end{minipage}
    	\begin{minipage}{0.34\linewidth}
		\centering
		\includegraphics[width=0.95\linewidth]{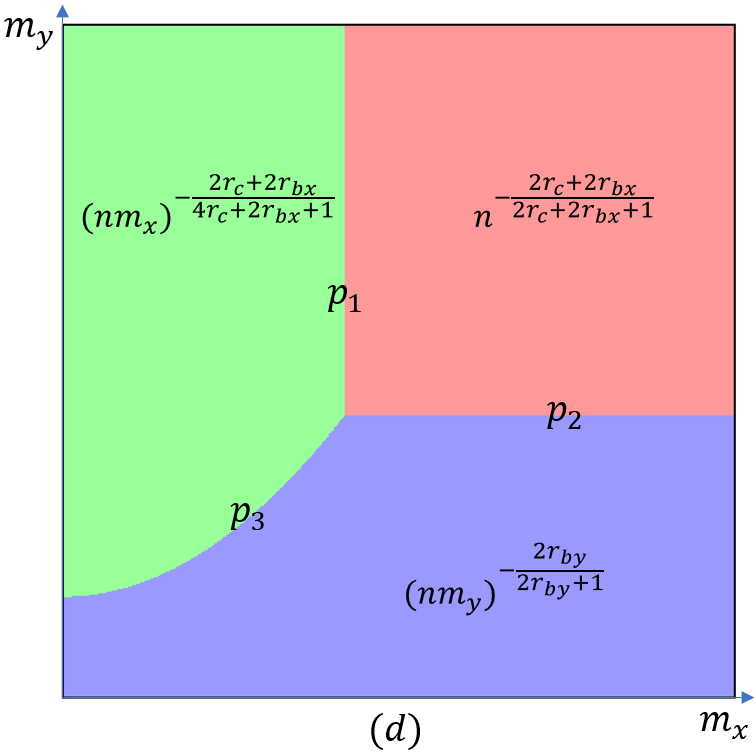}
	\end{minipage}
    \caption{ Phase diagram of function-on-function regression with discretely observed functional data. %
    The subfigure (a)-(d) correspond to the four cases: \(r_{bx} + r_c \leq r_{by}\), \(r_{by}< r_{bx} + r_c < r_{by}(1+2r_c)\), \( r_{bx} + r_c = r_{by}(1+2r_c)\) and \( r_{bx} + r_c > r_{by}(1+2r_c)\) respectively.
    The horizontal (vertical) coordinate represents the order of the predictor (response) sampling frequency with respect to \(n\), increasing further to the right (up).
    The boundaries \(p_1\), \(p_2\), \(p_3\) indicate phase transition positions in \eqref{equ:phasetransitions}, respectively. 
    They divide the convergence into several sampling regimes; where 
    red indicates the dense regime, while green/blue parts are the predictor/response-sparse regime.
    }\label{fig:phase}
\end{figure}

In the illustrative case for function-on-function regression, we assume that the slope function \( \bo \) belongs to a tensor product RKHS, \( \mathcal{H}(K) = \mathcal{H}(K^y) \otimes \mathcal{H}(K^x) \), where \( K = K^y \otimes K^x \) with \( K^y \) and \( K^x \) being kernels defined on the domains of \( Y \) and \( X \), respectively. Additionally, we assume that the simple eigenvalue decay rate of \(K^x\), \(K^y\) and \(C\) are \(k^{-2r_{bx}}\), \(k^{-2r_{by}}\) and \(k^{-2r_c}\), respectively; while \(C\) is assumed to share the aligned eigenfunctions with \(K^x\). Intuitively, \(r_c\) is the  ``smoothness'' of predictor functions, and \(r_{bx},r_{by}\) are ``marginal smoothness'' of the slope function \( \bo \) with respect to the predictor and response.\label{rcrbfirst}
Then, the minimax optimal rate of function-on-function regression is shown to be 
\[
n^{-\frac{2r_{bx} + 2r_c }{2r_{bx} + 2r_c  + 1}} + (nm_x)^{-\frac{2r_{bx} + 2r_c }{2r_{bx} + 4r_c + 1}} + (nm_y)^{-\frac{2r_{by}}{2r_{by} + 1}},
\]
where the first term reflects the influence of sample size \(n\), consistent with prior results for fully observed data such as \citet{lian2015minimax}; the second/third term captures the effect of discretization of the predictor/response functions, first revealed by this work; as illustrated in Figure \ref{fig:shiyi}. 
If \(r_{bx} + r_c \leq r_{by}\), the response functions are very smooth so that the influence of their discretization is small. 
Therefore, as shown in Figure \ref{fig:phase}\((a)\), the convergence phase behavior is similar to that in scalar-on-function regression. 
If \(r_{bx} + r_c > r_{by} \), the phase transition phenomena become more intricate, occurring at 
\begin{equation}\label{equ:phasetransitions}
    m_x \asymp n^{\frac{2r_c}{2r_{bx}+2r_c+1}}, \  m_y \asymp n^{\frac{r_{bx}+r_c-r_{by}}{(2r_{bx}+2r_c+1)r_{by}}} \  \text{and} \  m_y \asymp n^{\frac{r_{bx}+r_c-r_{by}-2r_c r_{by}}{(2r_{bx}+4r_c+1)r_{by}}} m_x^{\frac{(r_{bx}+r_c)(2r_{by}+1)}{(2r_{bx}+4r_c+1)r_{by}}}.
\end{equation}
These three boundaries partition the convergence into three phases, corresponding to the dense sampling regime and predictor/response-sparse sampling regime.
As depicted in Figure \ref{fig:phase}\((b)\)-\((d)\), the boundary between the sparse regimes of predictor and response exhibits slightly distinct behavior depending on the different relative magnitudes of \( r_{bx} + r_c \) and \( r_{by}(1 + 2r_c) \).
Nonetheless, for \(r_{bx} + r_c > r_{by}\), when \(m_x \gtrsim n^{2r_c/(2r_{bx}+2r_c+1)}\) 
and \(m_y \gtrsim n^{(r_{bx}+r_c-r_{by})/((2r_{bx}+2r_c+1)r_{by})}\), 
both the predictor and response functions are densely observed, and the convergence rate is of order \(n^{-(2r_{bx} + 2r_c)/(2r_{bx} + 2r_c  + 1 )}\). 
On the other hand, when \(m_x \lesssim n^{2r_c/(2r_{bx}+2r_c+1)}\)
the predictors are considered as sparsely sampled; while \(m_y \lesssim n^{(r_{bx}+r_c-r_{by})/((2r_{bx}+2r_c+1)r_{by})}\) defines the sparse design of response functions. 
If both predictor and response are sparsely sampled, then \(m_y^{{(2r_{bx}+4r_c+1)r_{by}}} \gtrsim (\mbox{respectively}, \lesssim) n^{r_{bx}+r_c-r_{by}-2r_c r_{by}} m_x^{(r_{bx}+r_c)(2r_{by}+1)}\) indicates that the sparsity of predictor (\mbox{respectively}, response) dominates.

In summary, our contributions are twofold. 
Methodologically, we introduce new pooling-ridge estimators for discretely observed functional linear regression problems, whether the response variable is a scalar or a function. 
Utilizing unbiased operator estimation with the pooling strategy, the bias typically introduced by sparse observations in traditional methods is overcome. 
These estimators are proven to achieve minimax optimality, which firstly addresses a long-standing open problem in FDA \citep{yao2005a,yao2005b,hall2007methodology}.
Theoretically, we provide a thorough understanding of the minimax optimal rates, fully characterizing the impact of discrete sampling in functional linear models. Notably, we accurately localize phase transition boundaries that have not been identified before. 
In scalar-on-function regression, the transition only divides the convergence into two distinct phases; while the function-on-function model may exhibit transitions at up to three phases. 
Additionally, numerical studies support the advantages of the proposed method.

The paper is organized as follows. In Section \ref{sec:method}, following an overview of preliminaries on functional data and RKHS, we introduce the proposed estimation methods for both scalar-on-function and function-on-function regression problems.
Then Section \ref{sec:theory} develops the statistical theory for the proposed estimators, addressing the minimax optimality, determining the phase transition boundaries, and analyzing some common cases. 
Finally, in Sections \ref{sec:simulation} and \ref{sec:realdata}, we present simulation experiments and real-world data applications, supporting the strengths of the proposed methods.%

\section{Model and methodology}\label{sec:method}
In this section, we first introduce some basic concepts related to functional data and RKHS. 
Following these preliminaries, the scalar-on-function and function-on-function regression models will be illustrated along with the proposed methods. 

\subsection{Preliminaries}\label{sec:prelim}
Consider a random function \( X(t) \) defined on the closed interval \(\mathcal{T}_x\), and 
let \( X_1(t), X_2(t), \dots, X_n(t) \) be its \( n \) independent and identically distributed (i.i.d.) realizations.
Throughout the paper, all integrals on \(\mathcal{T}_x\) and \(\mathcal{T}_y\) are taken with respect to the design measures \(\nux\) and \(\nuy\), respectively. Here, \(\nux\) and \(\nuy\) are assumed to have non-zero continuous densities for convenience.
The mean function \( \mu_X(t) = \mathbb{E}[X(t)] \) and the covariance function \( C(t_1, t_2) = \mathrm{Cov}(X(t_1), X(t_2)) \) are two basic concepts; we assume \(\sup_{t}\mathbb{E}[|X(t)|^4] \lesssim 1\). %
For simplicity, we assume that \( \mu_X(t) = 0 \) for all \( t \in \mathcal{T}_x \), as the estimation of the mean function has been extensively studied \citep{cai2011optimal,zhang2016sparse,yan2025drrm}.
By Mercer's theorem, the covariance \( C(t_1, t_2) \) admits the following spectral decomposition
\[
C(t_1, t_2) = \sum_{k=1}^\infty \mu_k \phi_k^C(t_1) \phi_k^C(t_2),
\]
where \( \mu_1 \geq \mu_2 \geq \cdots \geq 0 \) are the eigenvalues, and \( \phi_1^C, \phi_2^C, \ldots \) are the corresponding orthonormal eigenfunctions.  
In this work, we assume that \(C\) is positive definite; hence \(\mu_k > 0\) for each \(k\).
Each realization \( X_i(t) \) can then be expanded using the Karhunen–Loève expansion  
\(
X_i(t) = \sum_{k=1}^\infty \mu_k^{1/2} \xi_{ik} \phi_k^C(t),
\)  
where \( \mu_k^{1/2} \xi_{ik}\)'s are the principal component scores. The random variables \( {\xi_{ik}}\)'s, given by
\(
\xi_{ik} = \mu_k^{-1/2} \int_{\mathcal{T}_x} X_i(t) \phi_k^C(t) \, d\nux(t)
\)
are uncorrelated, with zero mean and unit variance. 
The decay rate of the eigenvalues \(\{\mu_k\}\) determines the smoothness of the covariance operator and the predictor process \(X\); faster decay corresponds to smoother realizations of \(X\), as high-frequency components are attenuated more rapidly.

Reproducing Kernel Hilbert Space (RKHS) plays a crucial role in nonparametric statistical analysis. Let \(\mathcal{T}\) be a compact set, and consider a reproducing kernel \(K: \mathcal{T} \times \mathcal{T} \rightarrow \mathbb{R}\) that is a real, symmetric, square-integrable and nonnegative definite function.   Each such kernel uniquely determines an RKHS, denoted \(\mathcal{H}(K)\), which is a Hilbert space with an associated inner product \(\langle \cdot, \cdot \rangle_{\mathcal{H}(K)}\).
A fundamental property is that the function \(K(u, \cdot)\) is an element of \(\mathcal{H}(K)\) for any \(u \in \mathcal{T}\). Moreover, the kernel exhibits the reproducing property
\[
f(u) = \langle K(u, \cdot), f \rangle_{\mathcal{H}(K)}, \quad \forall f \in \mathcal{H}(K).
\]  
The tensor product of RKHS is a commonly used concept. Given two domains, \(\mathcal{T}_y\) and \(\mathcal{T}_x\), with corresponding kernels \(K^y\) and \(K^x\), the tensor product kernel \(K = K^y \otimes K^x\) is defined by \(K(v_1,u_1;v_2,u_2) = K^y(v_1,v_2) K^x(u_1,u_2)\) accompanied with \(\mathcal{H}(K) = \mathcal{H}(K^y) \otimes \mathcal{H}(K^x) \). 
Let \(\{f_i: i \geq 1\}\) and \(\{g_j: j \geq 1\}\) be orthogonal basis sets for \(\mathcal{H}(K^y)\) and \(\mathcal{H}(K^x)\), respectively. Then, \(\{f_i g_j: i, j \geq 1\}\) forms an orthogonal basis set for the tensor product space \(\mathcal{H}(K^y) \otimes \mathcal{H}(K^x)\).
For more details on this topic, readers may refer to \citet{wahba1990spline}.

Building on the above concepts, we define the sampling operator \(K_u\) and its adjoint \(K_u^*\) as  
\[
K_u: \mathbb{R} \rightarrow \mathcal{H}(K), \, y \mapsto y K(u, \cdot),
\quad \text{and} \quad 
K_u^*: \mathcal{H}(K) \rightarrow \mathbb{R}, \, f \mapsto f(u),
\]  
for all \(u \in \mathcal{T}\).  
Moreover, we introduce  
\(S_K: \mathcal{L}^2(\mathcal{T},\nu) \rightarrow \mathcal{H}(K),%
f \mapsto \int_{\mathcal{T}} K(\cdot, u) f(u) \, d\nu(u). \)
It is well known that 
\(\int_{\mathcal{T}} f(t) \beta(t) \, d\nu(t)=\langle S_K f, \beta\rangle_{\mathcal{H}(K)}\) for any \(f \in \mathcal{L}^2(\mathcal{T},\nu)\) and \(\beta \in \mathcal{H}(K)\).
Thus the adjoint operator \(S_K^*\) of \(S_K\) is the natural embedding operator from \(\mathcal{H}(K)\) to \(\mathcal{L}^2(\mathcal{T},\nu)\). 
Finally, we define the integral operator
\[
L_K = S_K^* S_K: \mathcal{L}^2(\mathcal{T},\nu) \rightarrow \mathcal{L}^2(\mathcal{T},\nu),
f \mapsto \int_{\mathcal{T}} K(\cdot, u) f(u) \, d\nu(u). \]
Let \(\{(\lambda_k,\varphi_k^K):k\geq1\}\) denote the eigensystem of
\(L_K\). For \(\tau >0\), we define the fractional power \(L_K^\tau\) by
\(
L_K^\tau f
=
\sum_{k=1}^{\infty}
\lambda_k^\tau
\langle f,\varphi_k^K\rangle_{L^2}
\varphi_k^K,
\)
and, whenever well-defined, denote its associated kernel by
\(
K^{\tau}(t_1,t_2)
=
\sum_{k=1}^{\infty}
\lambda_k^\tau
\varphi_k^K(t_1)\varphi_k^K(t_2).
\)
Without special instructions, in the remaining of this paper, we will default to \(L_K\) as a positive definite, trace class operator, i.e. \(\operatorname{tr}(L_K) = \int_{\mathcal{T}}K(u,u)\,d\nu(u) < \infty\). We also assume that all reproducing kernels used later in the estimation procedures are continuous.

\subsection{Pooling-ridge estimation in scalar-on-function regression}\label{sec:modelsf}
Consider the following functional linear regression model with a scalar response 
\[
Y_i = \ao + \int_{\mathcal{T}_{x}} X_i(t) \bo(t) \, d\nux(t) + e_i, \quad 1\leq i \leq n,
\]
where \(\ao \in \mathbb{R}\) is the intercept, \(\bo\) is an unknown slope function defined on a compact set \(\mathcal{T}_{x}\) and \(e_i\) denotes independent random noise with zero mean and finite variance. For simplicity, we assume \(\alpha_0 = 0\) without loss of generality.

In practice, the trajectory of \( X_1(t), X_2(t), \dots, X_n(t) \) usually cannot be fully observed.
Instead, for the \(i\)-th individual, measurements are typically collected at \(m_{x,i}\) discrete time points, \(\{T_{i1}, T_{i2}, \ldots, T_{im_{x,i}}\}\), within the domain \(\mathcal{T}_x\), and are further contaminated by additive noise. 
To facilitate analysis, we assume that the observation times \(T_{ij}'s\) are independently sampled from the probability measure \(\nux\) on \(\mathcal{T}_x\).
Thus, the observation can be modeled as
\begin{equation}\label{equ:lisanx}
    X_{ij} = X_i(T_{ij}) + \varepsilon_{ij}, \quad 1 \leq i \leq n, \, 1 \leq j \leq m_{x,i},
\end{equation}
where \(\varepsilon_{ij}\) represents i.i.d. measurement error with zero mean and { uniformly bounded} fourth-order moment. These noise terms are assumed to be independent of all other random variables.  
For convenience, we assume \(m_{x,i} \asymp m_{x}\) for \(1\leq i\leq n\); without misunderstanding, we will abbreviate all \(m_{x,i}\) to \(m_x\). 
Observations within the same subject are typically dependent, while those across different subjects are independent; this clustered dependence structure is a key characteristic of such discretely observed functional data.
Commonly, we call that functional data are sparsely observed when \(m_x\) is relatively small, and conversely densely observed; the criteria for distinguishing sparse from dense vary with the considered problem.

A naive method to handle the discrete observations is to pre-smooth each curve and then apply previous works for fully observed functional data, but this introduces an unignorable bias when the measurements are sparse \citep{yao2005b}. 
Note that the true slope function \(\beta_{0}\) minimizes the population risk 
\begin{equation}\label{equ:200}
    \mathbb{E}\left[ \left(Y-\int_{\mathcal{T}_x} X(t) \beta(t) \, d\nux(t)\right)^2\right], 
\end{equation}
which is equal to \( \var(e) + \iint (\beta-\bo) (t_1) C(t_1, t_2)(\beta-\bo) (t_2) \, d\nux(t_1)\, d\nux(t_2) \). In practice, the mathematical expectation can be approximated by \(n\) samples \(\{(Y_i, X_i): 1\leq i \leq n\}\).
Although we could not obtain \(\int_{\mathcal{T}_x} X_i(t) \beta(t) \, d\nux(t)\), the empirical approximation \(m_x^{-1}\sum_{j=1}^{m_x} X_{ij}\beta(T_{ij})\) is available.
This suggests to minimize 
\begin{equation}\label{equ:201}
 \Tilde{L}(\beta) =  \frac{1}{n} \sum_{i=1}^{n} \left( Y_{i} - \frac{1}{m_x}\sum_{j=1}^{m_x}X_{ij}\beta(T_{ij}) \right)^2, 
\end{equation}
in an appropriate sieve or with a penalization term. 
However, due to the square terms \(X_{ij}^2\beta(T_{ij})^2\) in the expansion, when \(m_x\) is small and \(n\to \infty\), 
the difference between \eqref{equ:201} and \eqref{equ:200} convergences to
\[
 \frac{1}{m_x} 
 \left(
 \int_{\mathcal{T}_x}(C(t,t) + \var(\varepsilon)) \beta^2(t) \, d\nux(t) - 
 \iint_{\mathcal{T}_x\times\mathcal{T}_x}\beta(t_1) C(t_1, t_2)\beta(t_2) \, d\nux(t_1)\, d\nux(t_2)
 \right),
\]
which is not negligible. 
Therefore, the loss function involving \eqref{equ:201} is unsuitable for the discretely observed functional linear regression problem, especially with small sampling frequency. 
To accommodate arbitrary sampling scheme ranging from sparse to dense designs, we drop the criterion \eqref{equ:201}.

To address the challenge introduced by sparse sampling and motivate a unified estimation framework for various sampling schemes, recalling \eqref{equ:200}, when \(\bo \in \mathcal{H}(K)\)
\[
 \left(\iint K_{t_1} C(t_1,t_2) K^{*}_{t_2}\, d\nux(t_1)\, d\nux(t_2)\right) \bo = \mathbb{E} \left[\int_{\mathcal{T}_x} Y X(t) K(t, \cdot)  \, d\nux(t) \right],
\]
where the operator \(K^{*}_{t}: K^{*}_{t}f=f(t)\) and the function \(K_{t}: K_{t}(u)=K(t,u)\). 
This motivates the definition of the linear operator \[\T: \mathcal{H}(K) \to \mathcal{H}(K)\quad \text{as} \quad  \Gamma = \iint K_{t_1} C(t_1,t_2) K^{*}_{t_2}\, d\nux(t_1)\, d\nux(t_2),\] which satisfies \(\Gamma(f) = \iint C(t_1,t_2) f(t_2) K(t_1, \cdot)\, d\nux(t_1)\, d\nux(t_2)\). 
Applying the pooling strategy that uses all observations, let
\[
\Tnm = \frac{1}{n}  \sum_{i=1}^{n} \frac{1}{m_{x}(m_{x}-1)} \sum_{j_1\neq j_2} X_{ij_1}X_{ij_2}K_{T_{ij_1}} K^{*}_{T_{ij_2}}, 
\]
where \(\Tnm(f) = \frac{1}{n}  \sum_{i=1}^{n} \frac{1}{m_{x}(m_{x}-1)} \sum_{j_1\neq j_2}  X_{ij_1}X_{ij_2} f(T_{ij_2}) K(T_{ij_1}, \cdot) \in \operatorname{Span}\{ K(T_{ij},\cdot) , 1\leq i \leq n, 1\leq j \leq m_{x} \} \subseteq \mathcal{H}(K).\)
It is easy to verify that \(\Tnm\) is an unbiased estimator of \(\T\) %
whenever \(m_x \to \infty\) or remains finite.
{
Additionally, although removing bias, 
measurement error still contributes to the variability of \(\widehat{\Gamma}\). 
Indeed, in the variance calculation for quantities involving \(\hat{\Gamma}-\Gamma\), repeated sampling indices generate terms involving moments such as
\(\mathbb{E}\{(X(t)+\varepsilon)^2(X(t')+\varepsilon')^2\}\), depending on 
fourth moments of \(X\) and \(\varepsilon\), which are uniformly bounded by 
assumption. Hence measurement error affects only multiplicative constants in convergence upper bound.} \label{variancejieshi}
Then we define the estimator as
\begin{equation}\label{equ:estimatorsf}
    \hat{\beta} = 
(\Tnm + \lambda I)^{-1}\frac{1}{n}\sum_{i=1}^{n} Y_{i}\frac{1} {m_{x}}\sum_{j=1}^{m_{x}}X_{ij} K(T_{ij},\cdot) \ \in \mathcal{H}(K),
\end{equation}
where the real number \(\lambda > 0\) is a tuning parameter for regularization, and \(I: \mathcal{H}(K) \to \mathcal{H}(K)\) is the identity operator. 
Notably, this estimator also works when the true slope function \(\bo \notin \mathcal{H}(K)\). 
With slight abuse of notation, we may also write \(\hat{\beta}\) as its natural embedding into \(\Ltwo\) in the remainder of this paper.

Note that the estimator \(\hat{\beta}\) lies in \(\operatorname{Span}\{ K(T_{ij},\cdot) , 1\leq i \leq n, 1\leq j \leq m_{x} \}\), regardless of whether the true slope function \(\bo\) belongs to \(\mathcal{H}(K)\).
To implement the proposed estimator, we assume that \(\hat{\beta} = \sum_{i=1}^{n} \sum_{j=1}^{m_{x}} c_{ij} K(T_{ij},\cdot)\) and solve for the coefficients from
\begin{equation}\label{equ:sfj}
(\Tnm + \lambda I )  \sum_{i=1}^{n} \sum_{j=1}^{m_{x}} c_{ij}   K(T_{ij},\cdot) = \frac{1}{n} \sum_{i=1}^{n} Y_{i}\frac{1}{m_{x}}\sum_{j=1}^{m_{x}}X_{ij} K(T_{ij},\cdot),
\end{equation}
where \( \Tnm  K(T,\cdot) = \frac{1}{n}  \sum_{i=1}^{n} \frac{1}{m_{x}(m_{x}-1)} \sum_{j_1\neq j_2} X_{ij_1}X_{ij_2}K(T,T_{ij_2})K(T_{ij_1}, \cdot )\).

\subsection{Pooling-ridge estimation in function-on-function regression}\label{sec:modelff}
Consider the functional linear regression model with functional response 
\begin{equation}\label{equ:211}
    Y_i(s) = \ao(s) + \int_{\mathcal{T}_{x}} X_i(t) \bo(s,t) \, d\nux(t) + U_i(s), 
\end{equation}
where \(\ao \in \Ltwo\) is the intercept function and \(\bo \in \mathcal{L}^2(\mathcal{T}_y \times \mathcal{T}_x,\nuy \otimes \nux)\) denotes the slope function.
Assume that \(U\) is a random process with zero mean and a continuous covariance function independent with other random variables, %
where \(U_i, 1\leq i \leq n\) are its i.i.d. realizations. 
We also assume that \(\ao =0\) for simplicity.
The predictor functions are observed according to \eqref{equ:lisanx}. 
Assume that the sample size is \(n\) and the sampling frequencies of the response functions are \(m_{y,i}\)'s; the observations of the  \(Y_i\)'s are generated as  
\begin{equation}\label{equ:212}
Y_{il} = Y_{i}(S_{il}) + e_{il}, \quad 1\leq i \leq n, 1\leq l \leq m_{y,i},
\end{equation}
where \(e_{il}\)'s are independent measurement errors with zero mean and bounded variance, and the observation points \(S_{il}\) are independently drawn from the probability measure \(\nuy\) on \(\mathcal{T}_{y}\).  
For convenience, we also assume that \(m_{y,i} \asymp m_{y}\) for some \(m_y\) and all \(1\leq i \leq n\); we will omit \(i\) in the subscript and abbreviate \(m_{y,i}\) as \(m_{y}\) without confusion.

To estimate the slope function in the model \eqref{equ:211} and \eqref{equ:212}, we still use the RKHS framework. 
Assume that \(K(s_1,t_1;s_2,t_2) = K^y(s_1,s_2)K^x(t_1,t_2)\) where \(K^y\) and \(K^x\) are kernel functions defined on \(\mathcal{T}_y \times \mathcal{T}_y\) and \(\mathcal{T}_x \times \mathcal{T}_x\), respectively. 
By the properties of tensor product, \(\mathcal{H}(K) = \mathcal{H}(K^y)\otimes\mathcal{H}(K^x)\). 
Similar to the scalar-on-function problem, we would not minimize
\begin{equation}\label{equ:201ff}
     \Tilde{L}(\beta) =  \frac{1}{n} \sum_{i=1}^{n} \frac{1}{m_y} \sum_{l=1}^{m_y} \left( Y_{il} - \frac{1}{m_x}\sum_{j=1}^{m_x}X_{ij}\beta(S_{il}, T_{ij}) \right)^2
\end{equation}
over a sieve set or with a penalization term, due to the significant bias when \(m_x\) is small. 
Note that the true \(\beta_{0}\) minimizes the risk 
\(
        \mathbb{E}\left[ \int_{\mathcal{T}_{y}} \left(Y(s)-\int_{\mathcal{T}_x} X(t) \beta(s,t) \, d\nux(t)\right)^2 \, d\nuy(s)\right]%
\), which implies 
\[
\left(\iiint K_{s,t_1} C(t_1,t_2) K^{*}_{s,t_2}\, d\nuy(s)\, d\nux(t_1)\, d\nux(t_2)\right) \bo = \mathbb{E} \left[\iint Y(s) X(t) K(s,t; \cdot,\cdot) \, d\nuy(s)\, d\nux(t) \right],
\]
where, for \(f\in \mathcal{H}(K)\), the operator \(K^{*}_{s,t}: K^{*}_{s,t} (f) = f(s,t)\); and for \(u\in\mathcal{T}_x, v\in \mathcal{T}_y\), the function \(K_{s,t}:K_{s,t}(v,u)=K(s,t;v,u)\). 
Denote \(\Gamma = \iiint K_{s,t_1} C(t_1,t_2) K^{*}_{s,t_2}\, d\nuy(s)\, d\nux(t_1)\, d\nux(t_2)\), then we define its unbiased estimator as
\[
\Tnm = \frac{1}{n} \sum_{i=1}^{n} \frac{1}{m_{y}m_{x}(m_{x}-1)} \sum_{l=1}^{m_{y}} \sum_{j_1\neq j_2} X_{ij_1}X_{ij_2}K_{S_{il},T_{ij_1}} K^{*}_{S_{il},T_{ij_2}}. 
\]
Subsequently, taking some positive number \(\lambda\) and the identity operator \(I: \mathcal{H}(K) \to \mathcal{H}(K)\), we define
\begin{equation}\label{equ:estimatorff}
    \hat{\beta} = (\Tnm + \lambda I )^{-1} \frac{1}{n} \sum_{i=1}^{n} \frac{1}{m_{x} } \sum_{j=1}^{m_{x}} \frac{1}{ m_{y}} \sum_{l=1}^{m_{y}} Y_{il} X_{ij} K(S_{il},T_{ij}; \cdot, \cdot), 
\end{equation}
which lies in \(\operatorname{Span}\{ K(S_{il},T_{ij}; \cdot, \cdot) , 1\leq i \leq n, 1\leq j \leq m_{x}, 1\leq l \leq m_{y}\}. \)
Hereafter, \(\hat{\beta}\) also denotes its natural embedding into \(\mathcal{L}^2(\mathcal{T}_y \times \mathcal{T}_x,\nuy \otimes \nux)\). 
To implement the estimation, similar to the scalar-on-function regression, we set \(\hat{\beta} = \sum_{i=1}^{n} \sum_{j=1}^{m_{x}} \sum_{l=1}^{m_{y}}  c_{ijl} K(S_{il},T_{ij}; \cdot, \cdot)\) and 
solve the equation
\begin{equation}\label{equ:ffj}
    (\Tnm + \lambda I ) \sum_{i=1}^{n} \sum_{j=1}^{m_{x}} \sum_{l=1}^{m_{y}}  c_{ijl}  K(S_{il},T_{ij}; \cdot, \cdot) =  \frac{1}{n} \sum_{i=1}^{n} \frac{1}{m_{x} } \sum_{j=1}^{m_{x}} \frac{1}{ m_{y}} \sum_{l=1}^{m_{y}}  Y_{il} X_{ij} K(S_{il},T_{ij}; \cdot, \cdot). 
\end{equation}

\begin{remark}
We close this section by reviewing relevant literature and highlighting the novelty of the proposed method. 
Most existing RKHS-based approaches (e.g., \citealp{yuan2010reproducing, sun2018optimal, sang2022nonlinear}) are developed for fully observed functional data or rely on recovery procedures, which do not work under sparse sampling designs. Alternatively, FPCA-based methods (e.g., \citealp{yao2005b, zhou2023functional}) pool observations across subjects to estimate principal components, thereby improving estimation efficiency for sparsely observed samples, but they typically involve more theoretical obstacles when analyzing the estimated eigenfunctions. Thus, we organically combine the advantages of both to accommodate a wide range of sampling frequencies. Instead of directly estimating the covariance function, we pool discrete measurements from all subjects to construct unbiased estimators of the operators appearing in RKHS-based formulations, which are then plugged in to obtain the final estimator. This yields a new unified framework applicable to both scalar-on-function and function-on-function regression models under arbitrary realistic sampling schemes.\label{fangfaxin}
\end{remark}

\section{Theoretical results: minimax optimality and phase transition}\label{sec:theory}
In this section, we establish theoretical guarantees for proposed pooling-ridge estimations of both scalar-on-function and function-on-function regression problems. For each case, we obtain the upper bound on convergence rate and the matching information lower bound, evidencing the minimax optimality. These analyses reveal some essential properties of functional linear models with discrete observations, especially the phase transition phenomena. 

Before presenting the detailed analysis, we introduce a basic assumption regarding the predictor distribution, which will be used in both the following two subsections.

\begin{assumption}\label{ass:conc}
There is a universal constant \(C\), such that, for any \(f \in {\mathcal{L}^2(\mathcal{T}_x,\nux)}\),
    \[
\mathbb{E}\left[\left(\int_{\mathcal{T}_x} X(t) f(t) {\, d\nux(t)}\right)^4 \right]\leq C\left(\mathbb{E}\left[\left(\int_{\mathcal{T}_x} X(t) f(t) {\, d\nux(t)}\right)^2\right]\right)^2.
\]
\end{assumption}

This assumption is widely employed in the existing literature on functional linear regression \citep{yuan2010reproducing,cai2012minimax,du2014penalized}. Essentially, it requires that the fourth moments of \( X \) exhibit sufficient concentration. In the case where \( X \) follows a Gaussian process, the condition is satisfied with \( C = 3 \), since the random variable \( \int_{\mathcal{T}_x} X(t) f(t) {\, d\nux(t)} \) is also Gaussian.

\subsection{Scalar-on-function regression}\label{sec:sft}
To evaluate the accuracy of an estimator \(\hat{\beta}\) in the scalar-on-function regression model, naturally embedded in {\(\mathcal{L}^2(\mathcal{T}_x,\nux)\)}, we consider the following squared error norm 
\[
\mathcal{E}_{\bo}(\hat{\beta}) = \| L_{C}^{1/2}(\hat{\beta}-\bo)\|_{{\mathcal{L}^2(\mathcal{T}_x,\nux)}}^2, 
\] 
where the integral operator \(L_C: {\mathcal{L}^2(\mathcal{T}_x,\nux)} \to {\mathcal{L}^2(\mathcal{T}_x,\nux)}, f \mapsto \int_{\mathcal{T}_x} C(\cdot,t)f(t){\, d\nux(t)}\). 
This norm quantifies the discrepancy between \(\hat{\beta}\) and the underlying true \(\bo\) in the sense of the predicted response variable \citep{cai2012minimax}, since
\(
\mathcal{E}_{\bo}(\hat{\beta}) = \mathbb{E}\big[\big(
 \int_{\mathcal{T}_{x}} X(t) \bo(t) {\, d\nux(t)} -  \int_{\mathcal{T}_{x}} X(t) \hat{\beta}(t) {\, d\nux(t)} \big)^2
\big], 
\)
where the expectation is taken with respect to a new realization of \(X(t)\).

Let \(\{\lambda_k:k\geq 1\}\) denote the eigenvalues of \(L_K\), arranged in non-increasing order, we assume \(\lambda_k\asymp k^{-2r_b}\) for some \(r_b>1/2\).
For theoretical convenience,
we now introduce the conjugations of \(\T\) and \(\Tnm\) on {\(\mathcal{L}^2(\mathcal{T}_x,\nux) \to \mathcal{L}^2(\mathcal{T}_x,\nux)\)} as \[\Pi= 
L_{K}^{-1/2}S_K^* \T S_K^{*-1} L_{K}^{1/2} \quad\text{ and }\quad \Pinm = L_{K}^{-1/2}S_K^* \Tnm S_K^{*-1} L_{K}^{1/2} , \]
where \(S_K^*: \mathcal{H}(K) \to {\mathcal{L}^2(\mathcal{T}_x,\nux)}\) denotes the natural embedding operator and 
\[
L_K: {\mathcal{L}^2(\mathcal{T}_x,\nux)} \rightarrow {\mathcal{L}^2(\mathcal{T}_x,\nux)},
\quad
f \mapsto \int_{\mathcal{T}_x} K(\cdot, t) f(t) {\, d\nux(t)}
\]
is defined in Section \ref{sec:prelim}.
{Since operator \(\Pi\) is the conjugate representation of \(\Gamma\) on \(\mathcal L^2\), they share the same nonzero spectrum. Hence, the eigenvalue decay of \(\Pi\) is exactly the eigenvalue decay of \(\Gamma\). We formulate the assumptions in terms of \(\Pi\) because this representation is more convenient for comparison with the existing RKHS literature and naturally accommodates the possibility that \(\beta_0\notin \mathcal H(K)\).}\label{gomnge} 
Assume that the eigenvalues of \(\Pi = L_{K}^{1/2} L_C L_{K}^{1/2}\) are \( \gamma_1\geq \gamma_2 \geq ... > 0 \) and the corresponding eigenfunctions are \( \{ \varphi_k: k\geq 1 \} \), it follows
\[
 \quad \Pi(\varphi_k) = \gamma_k \varphi_k, \quad \forall k\geq 1. 
\]
From the definition, the performance of the eigen-decomposition of \(\Pi\) is
determined by both kernel \( K \) and covariance \( C \), with their alignment also plays a crucial role. Importantly, the eigenstructure of \(\Pi\) fundamentally determines the convergence rate of the estimation procedure. To formalize these key characteristics, we introduce the following assumption.

\begin{assumption}\label{ass:model}
(a). The eigenvalues \( \{ \gamma_k: k\geq 1  \} \) of \(\Pi\) satisfy \(\gamma_k \asymp k^{-2r}\), and there exists a
corresponding orthonormal eigenbasis  \( \{ \varphi_k: k\geq 1  \} \) satisfy \(\gamma_k^{-1}\| L_K^{1/2} \varphi_k \|_{{\Ltwo}}^{2} \asymp k^{2s}\) for some \(r,s\)
such that \(1/2<r-s\leq r_b\), \(r-r_b>1/2\) and
the function \(K^{(r-r_b)/r_b}\) is bounded.

(b). Furthermore, we assume that the slope function \(\bo\) satisfies \(\|\Pi^{\alpha/2}L_{K}^{-1/2} \bo\|_{{\Ltwo}}^{2}  \lesssim 1 \) for some \(\alpha\): \(-1 \leq \alpha < (2r-2s-1)/(2r)\). 
\end{assumption}

{
This condition captures the properties of the covariance operator \(C\), the reproducing kernel \(K\), and the true slope function \(\bo\).
Assumption \ref{ass:model}(a) imposes asymptotic conditions on \(\{\gamma_k:k\ge 1\}\) and 
\(
\gamma_k^{-1} \| L_K^{1/2} \varphi_k(t) \|_{{\Ltwo}}^2. 
\)
The decay rate of \(\gamma_k\) characterizes the complexity of
\(\Pi = L_{K^{1/2} C K^{1/2}}\).
Intuitively, it quantifies the smoothness of the inner product
\(\langle X,\bo\rangle\), and thus the intrinsic difficulty of the functional linear regression problem.
Previous studies, such as \citet{cai2012minimax}, show that the eigenvalue decay of \(\Pi\) determines the minimax optimal convergence rate for fully observed functional linear regression, namely \(n^{-2r/(2r+1)}\).
The quantity
\(
\gamma_k^{-1} \| L_K^{1/2} \varphi_k(t) \|_{{\Ltwo}}^2 = \| L_K^{1/2} \varphi_k(t) \|_{{\Ltwo}}^2 / \| L_{K^{1/2} C K^{1/2}}^{1/2} \varphi_k(t) \|_{{\Ltwo}}^2
\)
describes the complexity of \(C\), or equivalently the smoothness of the predictor \(X\), under the action of \(K\) and \(\varphi_k\)'s.
It therefore captures the difficulty induced by discrete sampling.

The quantities in Assumption \ref{ass:model}(a) depend on the smoothness of \(C\) and \(K\), as well as on their interaction relative to the considered eigenbasis.
Consider a simplified setting in which \(C\) and \(K\) share the same eigenfunctions, that is,
\(\phi_k^C=\phi_k^K=\varphi_k\) for all \(k\ge 1\).
If the eigenvalues \(\{\mu_k\}\) of \(C\) satisfy \(\mu_k\asymp k^{-2r_c}\) and the eigenvalues \(\{\lambda_k\}\) of \(K\) satisfy \(\lambda_k\asymp k^{-2r_b}\), then
\(
\gamma_k \asymp k^{-2(r_c+r_b)},
\) and 
\(
\gamma_k^{-1}\bigl\|L_K^{1/2}\varphi_k\bigr\|_{{\Ltwo}}^2
\asymp k^{2r_c}.
\)}
More generally, an admissible eigensystem of \(\Pi\) may involve mixtures of eigenfunctions of \(K\) from different spectral scales.
Intuitively, lower-frequency
components of \(K\) may be distributed across higher-frequency directions of \(\Pi\), such that \(\|L_K^{1/2}\varphi_k\|_{\Ltwo}^2\) can decay more slowly than the eigenvalues of \(K\), which may lead to
\(r-s<r_b\).
See the lower-bound construction for a quantitative example.

Note that the series \(\sum_{k=1}^{\infty} k^{-2(r-s)} \lesssim \sum_{k=1}^{\infty}\|L_K^{1 / 2} \varphi_k(t)\|_{\mathcal{L}^2}^2 \leq \operatorname{tr}\left(L_K\right)=\int_{\mathcal{T}_x} K(s, s) d s<\infty\), we obtain \(2(r-s)>1\).
On the other hand, by the Schur--Horn inequality, the
diagonal elements of a positive trace-class operator in any
orthonormal basis are majorized by its eigenvalues. Comparing the
corresponding tail sums therefore requires
\(
k^{1-2(r-s)}\gtrsim k^{1-2r_b},
\)
which yields \(r-s\leq r_b\).
Moreover, since \(L_C\) is trace class, \(\sum_{k=1}^{\infty}\mu_k<\infty\),
which implies \(\mu_k=o(k^{-1})\). Note that
\(
\gamma_{2k-1}\leq \mu_k\lambda_k,
\)
\(\lambda_k\asymp k^{-2r_b}\) and
\(\gamma_k\asymp k^{-2r}\), it follows that
\(
k^{-2r}
\asymp \gamma_{2k-1}
\leq \mu_k\lambda_k
=o\!\left(k^{-2r_b-1}\right).
\)
Hence,
\(r-r_b>1/2.\)
Therefore, the conditions \(1/2< r-s \leq r_b\) and \(r-r_b>1/2\) are necessary.
The boundedness condition of
\(
K^{(r-r_b)/r_b}(t_1,t_2)
=
\sum_{k=1}^{\infty}
\lambda_k^{(r-r_b)/r_b}
\varphi_k^K(t_1)\varphi_k^K(t_2)
\)
is used for lower bound. Intuitively, it ensures uniformly bounded pointwise
variance for a predictor process whose covariance operator is aligned
with \(K\) and whose eigenvalues decay at least as fast as
\(k^{-2(r-r_b)}\).

In Assumption \ref{ass:model}(b), the condition \(\Pi^{\alpha/2}L_{K}^{-1/2} \bo \in {\mathcal{L}^2(\mathcal{T}_x,\nux)}\) characterizes the smoothness of the slope function; allowing a model misspecification where \(\bo \notin \mathcal{H}(K)\) or with improved regularity that enable faster convergence rates.
Specifically, when \(\alpha = 0\), it implies \(\bo \in \mathcal{H}(K)\), while larger values of \(\alpha\) correspond to lower regularity of \(\bo\). 
The lower bound \(\alpha \geq -1\) is standard for kernel ridge regression \citep{li2024saturation}.
To provide intuition for the upper bound \(\alpha < (2r - 2s - 1)/(2r)\), consider the toy case where \(C\) and \(K\) share the same eigenfunctions. Under the alignment condition, the upper bound \(\alpha < (2r - 2s - 1)/(2r)\) is equivalent to 
\(
\|\Pi^{-\alpha/2} L_K^{1/2}\|_{ \operatorname{HS}} < \infty,
\)
which is used to guarantee the compactness of the operator \( \Pi^{-\alpha/2} L_K^{1/2} \).

Although the operator \(\T\) is positive definite, the unbiased estimation \(\Tnm\) does not necessarily preserve this property. To ensure the well-posedness for the inverse of \(\Tnm + \lambda I\), we establish the following result by bounding the largest negative eigenvalue of \(\Tnm\) in absolute magnitude.

\begin{proposition}\label{prop:positive}
Under Assumption \ref{ass:conc} and \ref{ass:model}, if we take 
\[
\lambda \asymp n^{-\frac{2 r}{2(1-\alpha)r+1}} + (nm_x)^{-\frac{2 r}{2(1-\alpha)r+2s+1}} ,
\]
then \(\ppp\left( (\Tnm + \lambda I) \text{ is positive definite}  \right) \to 1\). 
\end{proposition}

From the proof of Proposition \ref{prop:positive}, the specification of the order of \(\lambda\) can be relaxed to \(\lambda^{-1} =o\left( n^{r} \wedge (nm_x^2)^{r/(2s+1)}\right)\)  where \(a\wedge b\) indicates \(\min(a,b)\). In fact, such a choice is only for optimal convergence in the later analysis.
Now we are ready to present the main results of the convergence. 
First, the following result provides the upper bound of the convergence rate for the estimator \(\hat{\beta}\), with the appropriately chosen tuning parameter \(\lambda\).

\begin{theorem}\label{thm:uppersf}
Consider the scalar-on-function regression problem with discretely observed functional data. Under Assumption \ref{ass:conc} and \ref{ass:model}, let  
\[\lambda \asymp n^{-\frac{2 r}{2(1-\alpha)r+1}} + (nm_x)^{-\frac{2 r}{2(1-\alpha)r+2s+1}} ,\] then the estimator \(\hat{\beta}\) defined by \eqref{equ:estimatorsf} satisfies
\[
\mathcal{E}_{\bo}(\hat{\beta}) = O_p\left(n^{-\frac{2(1-\alpha)r}{2(1-\alpha)r+1}} + (nm_x)^{-\frac{2(1-\alpha)r}{2(1-\alpha)r+2s+1}}\right). 
\]
\end{theorem}

This result derives the convergence rate of the proposed estimator, which reflects the effect of sample size \( n \), sampling frequency \(m_x\), and the complexity parameters \( r,s \). 
{In the proof, we show the bias is \(O_p\big(\lambda^{1 - \alpha}\big)\) and the variance is \(O_p\big(n^{-1}\lambda^{-1/2r}    + (n m_x)^{-1} \lambda^{- (2s+1)/2r}\big)\); the tuning parameter \(\lambda\) balances this trade-off.}
To show that this rate is optimal, we also derive a matching information lower bound. 
Let \(\mathcal{P}\) denote the probability measure of the random variables that satisfy the basic setting of the scalar-on-function regression model in Section \ref{sec:modelsf}. %
Next, we proceed to establish an information lower bound for all estimators based on discretely observed data.

\begin{theorem}[Lower bound]\label{thm:lowbsf}
    Consider the scalar-on-function regression problem with discretely observed functional data. There is a constant \(c>0\) independent of \(n\) and \(m_x\) such that 
  \[
\varliminf_{n\to \infty} \inf_{\tilde{\beta}} \sup_{\mathcal{P} \in \mathcal{Q}}
\ppp\left(
\mathcal{E}_{\bo}(\tilde{\beta}) > c (n^{-\frac{2(1-\alpha)r}{2(1-\alpha)r+1}} + (nm_x)^{-\frac{2(1-\alpha)r}{2(1-\alpha)r+2s+1}}) 
\right) >0. 
\]  
where the infimum is taken over all possible estimators \(\tilde{\beta}\) based on the observation data set \(\left\{ \left(Y_{i},X_{ij},T_{ij} \right): 1\leq i \leq n, 1\leq j \leq m_x \right\}\), 
the supremum is taken over all distributions \(\mathcal{P} \in \mathcal{Q}\), where \(\mathcal{Q}\) is a collection of \(\mathcal{P}\) that further satisfy Assumptions \ref{ass:conc} and \ref{ass:model}. %
\end{theorem}

The upper bound derived in Theorem \ref{thm:uppersf} and the lower bound established in Theorem \ref{thm:lowbsf} coincide, confirming that the minimax optimal convergence rate for discretely observed scalar-on-function regression is \(n^{-2(1-\alpha)r/(2(1-\alpha)r+1)} + (nm_x)^{-2(1-\alpha)r/(2(1-\alpha)r+2s+1)}\). 
Notably, this result reveals a phase transition phenomenon in the minimax convergence rate. Specifically, when the sampling frequency satisfies \(m_x \asymp n^{2s/(2(1-\alpha)r+1)}\), the convergence rate reaches \(n^{-2(1-\alpha)r/(2(1-\alpha)r+1)}\), which depends only on the sample size \(n\). Beyond this boundary, increasing the number of repeated measurements \(m_x\) yields no further improvement in the rate, even in the ideal scenario where functions are fully observed \citep{yuan2010reproducing, cai2012minimax,dou2012estimation, du2014penalized,qu2016optimal}. 
On the other hand, in the sparse sampling regime, which is defined by \(m_x \lesssim n^{2s/(2(1-\alpha)r+1)}\), the convergence rate shifts to \((nm_x)^{-2(1-\alpha)r/(2(1-\alpha)r+2s+1)}\). This term highlights the effect of predictor discrete sampling, which has not been reported in previous studies. 
The above results completely characterize the scalar-on-function regression with discretely observed functional data, which is essentially an infinite-dimensional inverse problem confounded by discrete and noisy observations.

After establishing the optimal convergence rate, we now revisit two common special cases. 
In longitudinal data studies, the number of repeated measurements \(m_x\) for each subject is often finite, which is a typical application of functional data analysis \citep{hsing2015theoretical,wang2016functional}. 
Under the same conditions, the minimax optimal rate is of order 
\(
n^{-2(1-\alpha)r/(2(1-\alpha)r+2s+1)},
\)
which can be achieved by the proposed estimator. %
Even if there is no need to account for the asymptotic behavior of sampling frequency, this convergence rate has not been found elsewhere.
Another issue is that most studies assume the model is well-specified, i.e. \(\bo\) exactly belongs to \(\mathcal{H}(K)\); hence the results with \(\alpha = 0\) might be more widely used.

\begin{remark}\label{rmk:algsf}
The preceding rates are determined by the parameters \(r\) and \(s\), which characterize the complexity of the kernel \(K\), the covariance function \(C\), and their alignment. 
Now we can consider a toy case where \(K\) and \(C\) share aligned eigenfunctions. Although this assumption may not reflect the actual situation, it serves to provide an intuitive understanding.
Strictly speaking, assume that the eigenvalues \(\mu_k\)'s of \(C\) satisfy \(\mu_k \asymp k^{-2r_c}\) and the eigenvalues \(\lambda_k\)'s of \(K\) satisfy \(\lambda_k \asymp k^{-2r_b}\), with aligned eigenfunctions \(\phi_{k}^C=\phi_{k}^K\) for all \(k\geq 1\).
Under Assumption \ref{ass:conc} and assume that \(\bo \in \mathcal{H}(K)\), let  
\[\lambda \asymp n^{-\frac{2r_b + 2r_c }{2r_b + 2r_c  + 1}} + (nm_x)^{-\frac{2r_b + 2r_c }{2r_b + 4r_c  + 1}} ,\] then the estimator \(\hat{\beta}\) defined by \eqref{equ:estimatorsf} satisfies
\[
\mathcal{E}_{\bo}(\hat{\beta}) = O_p\left(n^{-\frac{2r_b + 2r_c }{2r_b + 2r_c + 1}} + (nm_x)^{-\frac{2r_b + 2r_c }{2r_b + 4r_c  + 1}}\right), 
\]
{and
\[
\|\hat{\beta} - \bo\|_{{\Ltwo}}^2 = O_p\left(n^{-\frac{2r_b}{2r_b + 2r_c  + 1}} + (nm_x)^{-\frac{2r_b }{2r_b + 4r_c  + 1}}\right). 
\]
The above results are minimax optimal. }
\end{remark}

{
Similar to \citet{yuan2010reproducing}, we establish the \({\mathcal{L}^2(\mathcal{T}_x,\nux)}\) estimation error under the alignment case.
Under this toy case, and for both two norms, the phase transition occurs at \( m_x \asymp n^{2r_c/(2r_b + 2r_c + 1)} \). 
Specifically, when the sampling frequency \( m_x \) exceeds this threshold, the convergence rate stabilizes at \(n^{-(2r_b + 2r_c)/(2r_b + 2r_c + 1)} \) for \(\mathcal{E}_{\bo}\) and \(n^{-2r_b /(2r_b + 2r_c + 1)} \) for \(\mathcal{L}^2\) estimation error; while under sparse sampling, the rate deteriorates to \( (nm_x)^{-(2r_b + 2r_c)/(2r_b + 4r_c + 1)} \) for \(\mathcal{E}_{\bo}\) and \( (nm_x)^{-2r_b/(2r_b + 4r_c + 1)} \) for \(\mathcal{L}^2\) estimation error.
Surprisingly, the phase transition boundary exhibits a different dependence on \( r_b \) and \( r_c \). 
The threshold \( n^{2r_c/(2r_b + 2r_c + 1)} \) decreases marginally as \( r_b \) increases but grows monotonically with \( r_c \). 
The former can be intuitively explained by the fact that a larger \( r_b \) reduces model complexity, leading to a faster convergence. 
However, while increasing \(r_c\) also reduces model complexity, it simultaneously diminishes the information available about higher-frequency features (e.g. eigenfunctions with large index) in the discretely observed functional data. 
The latter effect plays a crucial role in this interesting phenomenon: for large \( r_c \), more observations are required to adequately capture the required high-frequency information in the predictor, thereby delaying the phase transition, which we refer to as ``{\it curse of smoothness}''. This phenomenon does not occur in simple problems like mean and covariance estimation, but arises specifically in the functional linear model as it is essentially an infinite-dimensional inverse problem. }\label{cureseofsm}

To conclude this subsection, we provide a comparison with existing results in Table \ref{tab:ressf}. Previous studies have mainly focused on fully observed data \citep{hall2007methodology, yuan2010reproducing, dou2012estimation}. \citet{zhou2023functional} considered discrete observations and established convergence rates under a dense design. However, their analysis did not precisely identify the phase transition, resulting in a higher requirement of sampling frequency than necessary for the dense regime.  
In summary, our primary theoretical contributions are twofold. First, we accurately characterize the phase transition, refining the previous definition of dense design. Consequently, the required sampling frequency to achieve the same convergence rate as the fully observed case is significantly reduced. Second, we establish, for the first time, the minimax optimal rate under the sparse design, which has not been given in the previous works even in some special cases.

\begin{table}[ht]
  \caption{A table comparing newly established rates with existing results for scalar-on-function regression. } \label{tab:ressf}%
   \resizebox{0.6\textwidth}{!}{
  \begin{tabular}{cccc}
\hline
  &\rule[-1.8mm]{0pt}{6mm} {\large  Sparse} & {\large Phase transition} & {\large Dense} \\
 \hline
 {\large Proposed} & \rule{0mm}{6.8mm} %
 {\({(nm_x)^{-\frac{2r_b + 2r_c }{2r_b + 4r_c  + 1}}}\)} 
 & {  \({n^{ \frac{2r_c}{2r_b + 2r_c + 1} }} \)} & {  \({n^{-\frac{2r_b + 2r_c }{2r_b + 2r_c + 1}}}\)} \\
 {\large Existing} & ------ & \rule[-2mm]{0pt}{8.8mm}  {  \( n^{ \frac{4r_c+2}{2r_b + 2r_c + 1} }\)}\(\vee\){  \( n^{ \frac{4r_b+2r_c+1}{4r_b + 4r_c + 2} }
 \)} & \({  {n^{-\frac{2r_b + 2r_c }{2r_b + 2r_c + 1}}}}\) \\ 
 \hline
\end{tabular}
}
  \begin{tablenotes}    
\item[Note:] 
The results in the first row are established in this work and they are sharp. Here, ``Phase transition" indicates the order at which phase transition in the convergence rate occurs. "Sparse" refers to sampling frequencies \( m_x \) lower than the phase transition boundary, while ``Dense" pertains to those denser. 
\end{tablenotes}  
\end{table}%

\subsection{Function-on-function regression}

In this part, we investigate the theoretical performance of function-on-function regression. 
For \(\bo, \hat{\beta} \in {\mathcal{L}^2(\mathcal{T}_y \times \mathcal{T}_x,\nuy \otimes \nux)}\), the following metric is employed to assess the estimation performance
\[
\mathcal{E}_{\bo}(\hat{\beta}) = \| (I \otimes L_{C})^{1/2}(\hat{\beta}-\bo)\|_{{\mathcal{L}^2(\mathcal{T}_y \times \mathcal{T}_x,\nuy \otimes \nux)}}^2, 
\]
where \(I: {\Ltwo} \to {\Ltwo}\) is the identity operator and \(L_C: {\Ltwo} \to {\Ltwo}, f \mapsto \int_{\mathcal{T}_x} C(\cdot,t)f(t){\,d\nux(t)}\) is the integral operator. 
It characterizes the discrepancy between the estimator \(\hat{\beta}\) and the true slope function \(\bo\), reflecting the prediction risk for the response variable \citep{sun2018optimal}. A straightforward derivation shows that this quantity can also be expressed as
\(
\mathcal{E}_{\bo}(\hat{\beta}) = \int_{\mathcal{T}_{y}} \mathbb{E}[(
 \int_{\mathcal{T}_{x}} X(t) \bo(s,t) {\,d\nux(t)} -  \int_{\mathcal{T}_{x}} X(t) \hat{\beta}(s,t) {\,d\nux(t)} )^2
] {\,d\nuy(s)},
\)
where the expectation is taken for a new realization of \(X(t)\).

We consider the tensor product Hilbert space and assume that the kernel function \(K(s_1,t_1;s_2,t_2)=K^{y}(s_1,s_2)K^{x}(t_1,t_2)\) where \(K^{y}\) and \(K^x\) are kernel functions defined on \(\mathcal{T}_y \times \mathcal{T}_y\) and \(\mathcal{T}_x \times \mathcal{T}_x\), respectively. 
Let \(\{\lambda_k^x:k\geq 1\}\) denote the eigenvalues of \(K^x\), arranged in non-increasing order, we assume \(\lambda_k^x\asymp k^{-2r_{bx}}\) for some \(r_{bx}>1/2\).
Analogous to the scalar-on-function regression problem, we consider the conjugations of \(\T\) and \(\Tnm\) in {\(\mathcal{L}^2(\mathcal{T}_y \times \mathcal{T}_x,\nuy \otimes \nux) \to \mathcal{L}^2(\mathcal{T}_y \times \mathcal{T}_x,\nuy \otimes \nux)\)} as \[\Pi= 
L_{K}^{-1/2}S_K^* \T S_K^{*-1} L_{K}^{1/2} \quad\text{ and }\quad \Pinm = L_{K}^{-1/2}S_K^* \Tnm S_K^{*-1} L_{K}^{1/2} , \]
where \(S_K^*: \mathcal{H}(K) \to {\mathcal{L}^2(\mathcal{T}_y \times \mathcal{T}_x,\nuy \otimes \nux)}\) is the natural embedding operator and
\[
L_K: {\Ltwo} \rightarrow {\Ltwo},
\quad
f \mapsto \int_{\mathcal{T}_y \times \mathcal{T}_x} K(\cdot,\cdot;s, t) f(s,t) {\,d\nuy(s)\,d\nux(t)}
\]
is defined in Section \ref{sec:prelim}.
By the tensor product structure, we denote \(\Piy:=L_{K^y}\) and \(\Pix:=L_{K^x}^{1/2}L_{C}L_{K^x}^{1/2}\), therefore
\[
\Pi = L_{K^y}\otimes (L_{K^x}^{1/2}L_{C}L_{K^x}^{1/2}) = \Piy \otimes \Pix. 
\]
Recall \(\Pix\) and \(\Piy\) are self-adjoint bounded linear operators, we assume that the eigenvalues of \(\Pix\) are ordered as \( \gamma_1^x\geq \gamma_2^x \geq ... \geq 0 \) and those of \(\Piy\) are \( \gamma_1^y\geq \gamma_2^y \geq ... \geq 0 \).
The corresponding eigenfunctions are denoted by \( \{ \varphi^x_u: u\geq 1 \} \) and \( \{ \varphi^y_v: v \geq 1 \} \), respectively.
For convenience, we denote \(\gamma_{uv} = \gamma^y_v \gamma^x_u\) and \(\varphi_{uv}= \varphi^y_v \otimes \varphi^x_u \) for any \(u,v \geq 1\).
It is clear that the eigenvalues \( \{ \gamma^x_u: u \geq 1 \} \) and eigenfunctions \( \{ \varphi^x_u: u \geq 1 \} \) depend on both \( K \) and \( C \), particularly their alignment.  
Additionally, the eigenvalues \( \{ \gamma^y_v: v \geq 1 \} \) characterize the marginal complexity of \(\Pi_y\) which determines the influence of the discrete observation of functional response.
We now introduce the following assumption.

\begin{assumption}\label{ass:modelff}
    (a). The eigenvalues \( \{ \gamma_u^x: u\geq 1 \} \) of \(\Pix\) satisfy \(\gamma_u^x \asymp u^{-2r_x}\) 
     and there exists an
     orthonormal eigenbasis
     \( \{ \varphi^x_u : u \geq 1\} \) satisfy \({(\gamma_u^{x})}^{-1}\| L_{K^x}^{1/2} \varphi_u^x(t) \|_{{\Ltwo}}^{2}  \asymp u^{2s}\) for some \(r_{x},s\)
    such that  \(1/2 <r_{x}- s\leq r_{bx}\) and \(r_x-r_{bx} > 1/2\) and
     \((K^x)^{(r_x-r_{bx})/r_{bx}}\) is bounded.
    
    (b). The eigenvalues \( \{ \gamma^y_v: v\geq 1 \} \) of \(\Piy\) satisfy \(\gamma_v^y \asymp v^{-2r_y}\). 
    Moreover, we assume that the functions \(\|\varphi^y_{v}\|_{\mathcal{L}^4} \lesssim 1\) uniformly.
    
    (c). Furthermore, we assume that \(\|\Pi^{\alpha/2}L_{K}^{-1/2} \bo\|_{{\Ltwo}}^{2}  \lesssim 1 \) for some \(\alpha \geq -1\) satisfying \(\alpha < (2r_x-2s-1)/(2r_x)\) and \(\alpha < (2r_y-1)/(2r_y)\).
 
\end{assumption}

In the above condition, (a) and (c) are similar to Assumption \ref{ass:model}. In term (b), we determine the marginal eigendecay rate of \(\Piy\) and assume that the eigenfunctions are uniformly bounded in the sense of \(\mathcal{L}^4\) norm, which is milder than the commonly assumed uniform boundedness in \(\mathcal{L}^\infty\) norm \citep{hsing2015theoretical}. 
Based on the previous assumption, we now demonstrate the asymptotic positive definiteness of \(\Tnm + \lambda I\), which guarantees that the proposed estimator is well-defined.

\begin{proposition}\label{prop:positiveff}
Under Assumption \ref{ass:conc} and \ref{ass:modelff}, if we take 
\[\lambda \asymp n^{-\frac{2 r_x}{2(1-\alpha)r_x+1}} + (nm_x)^{-\frac{2 r_x}{2(1-\alpha)r_x+2s+1}} + (nm_y)^{-\frac{2 r_y}{2(1-\alpha)r_y+1}} ,\]
then \(\ppp\left( (\Tnm + \lambda I) \text{ is positive definite}  \right) \to 1\). 
    
\end{proposition}

Similarly to Proposition \ref{prop:positive}, the order of the regularization parameter \(\lambda\) in Proposition \ref{prop:positiveff} is for statistical optimality in later analysis.
The following theorem bounds the convergence rate of the proposed estimator with \(\lambda\) at the appropriate order.

\begin{theorem}\label{thm:upperff}
Consider the function-on-function regression problem with discretely observed functional data. Under Assumption \ref{ass:conc} and \ref{ass:modelff}, let  
\[\lambda \asymp n^{-\frac{2 r_x}{2(1-\alpha)r_x+1}} + (nm_x)^{-\frac{2 r_x}{2(1-\alpha)r_x+2s+1}} + (nm_y)^{-\frac{2 r_y}{2(1-\alpha)r_y+1}} ,\] then the estimator \(\hat{\beta}\) defined by \eqref{equ:estimatorff} satisfies
\begin{equation}\label{equ:rateff}
    \mathcal{E}_{\bo}(\hat{\beta}) = O_p\left(n^{-\frac{2(1-\alpha)r_x}{2(1-\alpha)r_x+1}} + (nm_x)^{-\frac{2(1-\alpha)r_x}{2(1-\alpha)r_x+2s+1}} \log n + (nm_y)^{-\frac{2(1-\alpha)r_y}{2(1-\alpha)r_y+1}}  \log n \right), 
\end{equation}
where the first \(\log\) factor can be removed when \(r_x \neq (2s+1)r_y\) and the second \(\log\) factor can be removed when \(r_x \neq r_y\). 
\end{theorem}

{
In the proof, we show that the bias is \(O_p\big(\lambda^{1 - \alpha}\big)\) and the main term of variance is \(O_p\big( n^{-1} \lambda^{-1/(2r_x)} + (n m_x)^{-1} \lambda^{-(2s+1)/(2r_x)} \log (1/\lambda) + (nm_y)^{-1} \lambda^{ - 1/2r_y} \log (1/\lambda)  \big)\); hence the chosen \(\lambda\) balances their trade-off.} 
In distinction to Theorem \ref{thm:uppersf} for scalar-on-function regression problem, the result here introduces an additional term \((nm_y)^{-2(1-\alpha)r_y/(2(1-\alpha)r_y + 1)}\), which arises due to the discrete observations of the response functions.
Let \( \mathcal{P} \) denote the probability distribution of the random variables that adhere to the foundational settings of the function-on-function regression model, as described in Section \ref{sec:modelff}. We then define \( \mathcal{Q} \) as a set of \( \mathcal{P} \), constrained by some additional specific conditions. In the subsequent analysis, we derive a matching lower bound for all estimators based on discrete observations.

\begin{theorem}[Lower bound]\label{thm:lowbff}
    Consider the function-on-function regression problem with discretely observed functional data. There is a constant \(c>0\) independent of \(n\) and \(m_x,m_y\) such that
  \[
\varliminf_{n\to \infty} \inf_{\tilde{\beta}} \sup_{\mathcal{P} \in \mathcal{Q}}
\ppp\left(
\mathcal{E}_{\bo}(\tilde{\beta}) > c (n^{-\frac{2(1-\alpha)r_x}{2(1-\alpha)r_x+1}} + (nm_x)^{-\frac{2(1-\alpha)r_x}{2(1-\alpha)r_x+2s+1}} + (nm_y)^{-\frac{2(1-\alpha)r_y}{2(1-\alpha)r_y+1}}     ) 
\right) >0. 
\]  
where the infimum is taken over all possible estimators \(\tilde{\beta}\) based on the observed dataset \(\left\{ \left(Y_{il},S_{il},X_{ij},T_{ij} \right): 1\leq i \leq n, 1\leq j \leq m_x, 1\leq l \leq m_y   \right\}\), 
the supremum is taken over all distributions \(\mathcal{P} \in \mathcal{Q}\), where \(\mathcal{Q}\) is a collection of \(\mathcal{P}\) that further satisfy Assumptions \ref{ass:conc} and \ref{ass:modelff}. 
\end{theorem}

The matching results in Theorem \ref{thm:upperff} and Theorem \ref{thm:lowbff} determine the minimax optimal rate %
for the discretely observed function-on-function regression, ranging from sparse to dense design. %
The logarithmic factors can be removed when \(r_x \neq (2s+1)r_y\) and \(r_x \neq r_y\) and hence the optimality would be sharp. 
Notably, the second and third terms in the minimax rate reveal the effect of the discrete sampling of the predictor and response functions, respectively, which have not been reported in previous work.

Now, we discuss the behavior of the phase transitions. %
For the case where \(r_y \geq r_x\), the the slope surface is sufficiently regular in the response direction, and the influence of their discretization, i.e. the third term in \eqref{equ:rateff}, is necessarily dominated by the first term. 
Consequently, the phase transition only occurs at \(m_x \asymp n^{2s/(2(1-\alpha)r_x+1)}\). When the sampling frequency satisfies \(m_x \gtrsim n^{2s/(2(1-\alpha)r_x+1)}\), the convergence rate reaches \(n^{-2(1-\alpha)r_x/(2(1-\alpha)r_x+1)}\), and there would be no further improvement with larger \(m_x\) and \(m_y\). 
In contrast, in the sparse sampling regime, which is defined by \(m_x \lesssim n^{2s/(2(1-\alpha)r_x+1)}\), the convergence rate transitions to \((nm_x)^{-2(1-\alpha)r_x/(2(1-\alpha)r_x+2s+1)}\). 
Hence, it exhibits a phase transition behavior similar to that observed in scalar-on-function regression problems.

When \(r_y < r_x\), the situation becomes more complex. The phase transitions occur at %
\[
 m_x \asymp n^{\frac{2s}{2(1-\alpha)r_x + 1}} ,\  m_y \asymp n^{\frac{r_x - r_y}{(2(1-\alpha)r_x + 1)r_y}}  \text{ and } m_y^{} \asymp n^{
 \frac{r_x - r_y(2s + 1)}{r_y(2(1-\alpha)r_x + 2s + 1)}} 
 m_x^{ \frac{r_x(2(1-\alpha)r_y + 1)}{r_y(2(1-\alpha)r_x + 2s + 1)} } .
\]
Specifically, when the sampling frequencies satisfies \(m_x \gtrsim n^{2s/(2(1-\alpha)r_x+1)}\) and \(m_y \gtrsim n^{(r_x - r_y)/(2(1-\alpha)r_x + 1)r_y}\), the system enters the dense regime. Then the convergence rate is \(n^{-2(1-\alpha)r_x/(2(1-\alpha)r_x+1)}\), which only depends on the sample size \(n\). Beyond these thresholds, further increases in \(m_x\) and \(m_y\) do not improve the rate, even under the ideal condition of fully observed functions.
If \(m_x \lesssim n^{2s/(2(1-\alpha)r_x+1)}\), the predictor is considered to be sparsely sampled; if \(m_y \lesssim n^{(r_x - r_y)/(2(1-\alpha)r_x + 1)r_y}\), the response  is regarded as sparse. 
If the predictor is sparsely sampled but the response is dense, the convergence is in the predictor-sparse regime; and vice versa in the response-sparse regime. 
When both the predictor and response are sparse, the convergence behavior depends on which sparsity dominates. %
If \(m_y^{r_y(2(1-\alpha)r_x+2s+1)} \lesssim n^{(r_x - r_y(2s+1))} m_x^{r_x(2(1-\alpha)r_y + 1)}\), the sparsity of the response drives the dominant phase, and the corresponding convergence rate is \((nm_y)^{-2(1-\alpha)r_y/(2(1-\alpha)r_y + 1)}\). Conversely, %
the system would be in the predictor-sparse regime, %
with convergence rate of \((nm_x)^{-2(1-\alpha)r_x/(2(1-\alpha)r_x + 2s + 1)}\).
In summary, the phase behavior is presented in Table \ref{tab:resff}.

\begin{table}[htbp]
  \caption{A summary of the three convergence rate phases in function-on-function regression. All orders in this table are optimal and first established in this work for discretely observed functional data. } \label{tab:resff}%
 \resizebox{\textwidth}{!}{
  \begin{tabular}{ccc}
\hline
    &\rule[-1.8mm]{0pt}{6mm} {\large Phase position} & {\large Convergence rate} \\
 \hline
 {\large Dense regime} &\rule[0pt]{0pt}{6.8mm} {  \( m_x \gtrsim n^{\frac{2s}{2(1-\alpha)r_x + 1}} ,\  m_y \gtrsim n^{\frac{r_x - r_y}{(2(1-\alpha)r_x + 1)r_y}} \)} 
 & {\scriptsize\(n^{-\frac{2(1-\alpha)r_x}{2(1-\alpha)r_x+1}}\)} \\
 {\large Predictor-sparse regime} &\rule{0pt}{6.8mm} 
 {\scriptsize\( m_x \lesssim n^{\frac{2s}{2(1-\alpha)r_x + 1}}, m_y \gtrsim n^{
 \frac{r_x - r_y(2s + 1)}{r_y(2(1-\alpha)r_x + 2s + 1)}} 
 m_x^{ \frac{r_x(2(1-\alpha)r_y + 1)}{r_y(2(1-\alpha)r_x + 2s + 1)} } \)} & {   \( (nm_x)^{-\frac{2(1-\alpha)r_x}{2(1-\alpha)r_x+2s+1}}\)}\\
 {\large Response-sparse regime} &\rule[-2mm]{0pt}{8.8mm} {  \( \  m_y \lesssim n^{\frac{r_x - r_y}{(2(1-\alpha)r_x + 1)r_y}}, m_y \lesssim n^{
 \frac{r_x - r_y(2s + 1)}{r_y(2(1-\alpha)r_x + 2s + 1)}} 
 m_x^{ \frac{r_x(2(1-\alpha)r_y + 1)}{r_y(2(1-\alpha)r_x + 2s + 1)} } \)} & {  \( (nm_y)^{-\frac{2(1-\alpha)r_y}{2(1-\alpha)r_y+1}}\) }\\
 \hline
\end{tabular}
 }
  \begin{tablenotes}    
\item[Note:] The three rows in the table indicate three phases in convergence. Here, ``Phase position'' represents when the system is in this phase. Specifically, when \(r_x \leq r_y\), the response-sparse regime is empty.
\end{tablenotes}  
\end{table}

Another interesting finding is that when the sampling frequency \(m_y\) is sufficiently large, the convergence rate becomes independent of the marginal complexity of \(K^y\). Specifically, for fully observed data, the rate \(n^{-{2(1-\alpha)r_x}/({2(1-\alpha)r_x+1})}\) is determined only by the eigendecay of \(\Pix = L_{K^x}^{1/2} L_C L_{K^x}^{1/2}\), rather than that of \(\Pi\). In fact, note that \(\Pi = \Piy \otimes \Pix\), the eigendecay rate of \(\Pi\) will be slower than that of \(\Pix\) if \(r_y < r_x\).

The optimal convergence rate has been established; we now examine two common special cases. The first case arises when each subject is associated with finite sampling frequencies \(m_x\) and \(m_y\) \citep{yao2005b}. In this longitudinal scenario, the optimal rate is given by  
\(
n^{-2(1-\alpha)\tilde{r}/(2(1-\alpha)\tilde{r}+1)},\) where \(\tilde{r} = \min\{ r_x/(2s+1), r_y\}\).  
Another case corresponds to the well-specified model by setting \(\alpha = 0\), which is simpler and commonly considered in statistical theoretical works.

\begin{remark}\label{rmk:algff}
    Similar with Remark \ref{rmk:algsf}, we consider a toy example in which  \( C \) is aligned with \(K^x\) for intuitive illustration. %
    Assume that the eigenvalues \(\mu_k\)'s of \(C\) satisfy \(\mu_k \asymp k^{-2r_c}\) and the eigenvalues \(\lambda_u\)'s of \(K^x\) satisfy \(\lambda_u \asymp u^{-2r_{bx}}\), with aligned eigenfunctions \(\phi_{u}^C=\phi_{u}^{Kx}\) for any \(u\geq 1\). 
Suppose that Assumption \ref{ass:conc}, \ref{ass:modelff}(b) hold, and we denote \(r_{by}=r_{y}\) for symmetry of notations. 
Assume \(\bo \in \mathcal{H}(K)\), let  
\[\lambda \asymp n^{-\frac{2r_{bx} + 2r_c }{2r_{bx} + 2r_c  + 1}} + (nm_x)^{-\frac{2r_{bx} + 2r_c }{2r_{bx} + 4r_c + 1}} + (nm_y)^{-\frac{2r_{by}}{2r_{by} + 1}},\] then the estimator \(\hat{\beta}\) defined by \eqref{equ:estimatorff} satisfies
\[
\mathcal{E}_{\bo}(\hat{\beta}) = O_p\left(
n^{-\frac{2r_{bx} + 2r_c }{2r_{bx} + 2r_c  + 1}} + (nm_x)^{-\frac{2r_{bx} + 2r_c }{2r_{bx} + 4r_c + 1}} \log n + (nm_y)^{-\frac{2r_{by}}{2r_{by} + 1}} \log n
\right),
\]
where the first \(\log\) factor can be removed when \(r_x \neq (2s+1)r_y\) and the second \(\log\) factor can be removed when \(r_x \neq r_y\).  %
The asymptotic behavior has been illustrated in Section \ref{sec:motcon} and Figures \ref{fig:shiyi},\ref{fig:phase}; they divide the convergence phase into up to three parts, i.e. dense sampling, predictor sparse sampling dominance, and response sparse sampling dominance.
It is easy to see, similar to the scalar-on-function regression, the phase transitions also suffer a curse of the smoothness of predictor functions, since it is an infinite-dimensional inverse problem. 
{
Nevertheless, it is unclear whether the proposed estimator can achieve minimax optimal rate in terms of \(\mathcal{L}^2\) error. 
In fact, even for fully observed function-on-function regression \citep{lian2015minimax,sun2018optimal,dette2024statistical}, the optimality of RKHS estimator in terms of \(\mathcal{L}^2\) estimation error is still unestablished. 
Technically, the current analysis can bound quantities of the form
\(
\|\Pi^a L_K^{-1/2}(\hat{\beta} - \beta_0)\|_{\mathcal{L}^2}
\) for \(a \in [0,1/2]\) \citep[e.g. Theorem 6 in][]{yuan2010reproducing};  
but there is not an appropriate choice of \(a\) such that \(\|\Pi^a L_K^{-1/2}(\hat{\beta} - \beta_0)\|_{\mathcal{L}^2} \asymp \|\hat{\beta}-\beta_0\|_{\mathcal{L}^2}\) for function-on-function regression. We leave this problem for future investigation.}
\end{remark}

\section{Simulation}\label{sec:simulation}

In this section, to further demonstrate the practical merits of the proposed methodology, we conduct simulation studies.
Cross-validation is used for parameter tuning of the proposed method. 
Note when introducing the estimation procedures, we have dropped the empirical criteria \( \Tilde{L}(\beta)\) defined by \eqref{equ:201} and \eqref{equ:201ff}, due to the bias in sparse design. 
Since the theoretical optimality mainly benefits from the unbiasedness of \(\Tnm\), employing \(\Tilde{L}(\beta)\) for parameter selection would undermine the prior efforts.
Now, we introduce 
\(
\hat{L}(\beta) = \frac{1}{n } \sum_{i=1}^{n } ( Y_i^2 - 2Y_i \frac{1}{m_{x} } \sum_{j=1}^{m_x} X_{ij}\beta(T_{ij}) + \frac{1}{m_{x} (m_{x}-1)} \sum_{j_1\neq j_2} X_{ij_1}X_{ij_2}\beta(T_{ij_1})\beta(T_{ij_2}) )
\)
for scalar-on-function regression and its counterpart for functional response can be similarly defined. %
Despite \(\hat{L}(\beta)\) being unbiased for population risks, unfortunately, it is no longer nonnegative and may have large variance for ridiculous %
\(\beta\), hence leading to unstable parameter selection. 
To mitigate the bias and variance, we propose a two-step strategy that leverages both the stability of \(\Tilde{L}(\beta)\) and the unbiasedness of \(\hat{L}(\beta)\). The \(K\)-fold cross-validation computes the average performance \( \Tilde{L}(\beta_{\lambda})\) and \( \hat{L}(\beta_{\lambda})\) for each \(\lambda\) in the candidate parameter set \(\Lambda\).
First we make an initial selection using 
\(
\tilde{\Lambda} = \{ \lambda \in \Lambda : \tilde{L}(\beta_{\lambda}) \leq b \},
\)
where the threshold \( b \) is chosen, for instance, as  
\(
b = \alpha \min\{ \tilde{L}(\beta_{\lambda}) \mid \lambda \in \Lambda \}
\)
for some constant \( \alpha \geq 1 \),
which is taken to be \(1.2\) based on numerical experience. 
The choice can be adapted to the problem at hand.
In general, smaller values improve stability, whereas larger values retain a broader range of candidates.
Then, the tuning parameter is determined by   
\(
\hat{\lambda} = \arg \min_{\lambda \in \tilde{\Lambda}} \hat{L}(\beta_{\lambda}),
\)
which selects the candidate that minimizes the unbiased estimator \(\hat{L}\) within the subset \( \tilde{\Lambda} \). 
In the following, all cross-validation procedures are taken to be five-fold.

Without loss of generality, we set \(\mathcal{T}_{x}=\mathcal{T}_{y}=[0,1]\). 
Now we consider two designs for the predictor process. 
In the cosine design, we use the trigonometric basis \(\{\phi_k(t):k\geq 1\}\), where \(\phi_1(t)=1\) and \(\phi_k(t)=\sqrt{2}\cos((k-1)\pi t)\) for \(k\geq 2\). 
In the Legendre design, we generate the predictor process using the shifted Legendre polynomial basis \(\{\ell_k(t):k\geq 1\}\) on \([0,1]\).
Specifically, letting \(\psi_k=\phi_k\) for the cosine design and \(\psi_k=\ell_k\) for the Legendre design, the predictor functions are generated as
\[
X_i(t)=\sum_{k=1}^{\infty} k^{-1}\xi_{ik}\psi_k(t),
\]
where the coefficients \(\xi_{ik}\) are independently drawn from \(N(0,1)\).
Then the discrete observations for each function are generated as 
\[
X_{ij} = X_{i}(T_{ij}) + \varepsilon_{ij}, \quad 1\leq i \leq n, 1\leq j \leq m_{x},
\]
where the measurement noises \(\varepsilon_{ij}\)'s follow i.i.d. Gaussian distribution \(N(0,0.5^2)\) and the observation points \(T_{ij}\)'s are independently drawn from uniform distribution on \(\mathcal{T}_{x}\).

\subsection{Scalar-on-function regression}\label{sec:simulationsf}
For scalar-on-function regression, the true slope function \(\bo\) is defined as
\[
\bo(t) = \sum_{k=1}^{\infty} 4(-1)^{k+1} k^{-2} \phi_k(t),
\]
where \(\{\phi_k(t):k\geq 1\}\) is the trigonometric basis. 
Further, the response \(Y_i\)'s are generated as
\(
Y_{i} = \int_{\mathcal{T}_{x}} X_{i}(t) \bo(t) dt + e_i
\) for \( 1\leq i \leq n\)
with measurement noises \(e_{i}\)'s i.i.d. following \(N(0,0.5^2)\). 
Under the Legendre design, the predictor process is generated from a different basis from slope function, which provides an additional robustness check for the misalignment between \(C\) and \(\bo\).
To implement the proposed method, we employ the commonly used Mat\'ern \(3/2\) kernel, given by \(K(t_1,t_2) = (1+|t_1-t_2|)\exp(-|t_1-t_2|)\). 
It is kept fixed in both designs and is not constructed to match the generating basis of the predictor process.
In this subsection, we set the sample size to \(n=50,100,200\) and the sampling frequency to \(m_x = 3,6,10,20\). Additionally, separate \(500\) i.i.d. subjects are generated as the test set to assess \(\mathcal{E}_{\bo}(\hat{\beta})\) for the estimators.

To comprehensively evaluate the proposed approach, several approaches are compared, including the plug-in method \citep{hall2007methodology}, the approximated least-square methods based on FPCA, and the RKHS methods for fully observed data \citep{yuan2010reproducing}. 
Specifically,  in the approximated least-squares methods, the functional principal component scores are estimated using three techniques, namely the integral approximation method (IN), the conditional expectation estimator \citep[PACE,][]{yao2005a}, and Monte Carlo estimator \citep[MC,][]{zhou2023functional}. 
When implementing RKHS methods for fully observed data, the required integrations involving the function \(X_i\) are approximated through empirical estimation using the discrete observations. For example, \(\int_{\mathcal{T}_x} X_{i}(t)f(t)dt\) is approximated by \(m_x^{-1} \sum_{j=1}^{m_x} X_{ij}f(T_{ij})\). 
These competing methods select their tuning parameters via a standard cross-validation procedure as in \citet{zhou2023functional}, and the candidate parameters of all methods are taken to be \(\Lambda = \{4^{-5},4^{-4},\ldots,4^{-1}\}\).

\begin{figure}[htbp]
\centering
\begin{minipage}{0.49\linewidth}
\centering
\includegraphics[width=\linewidth]{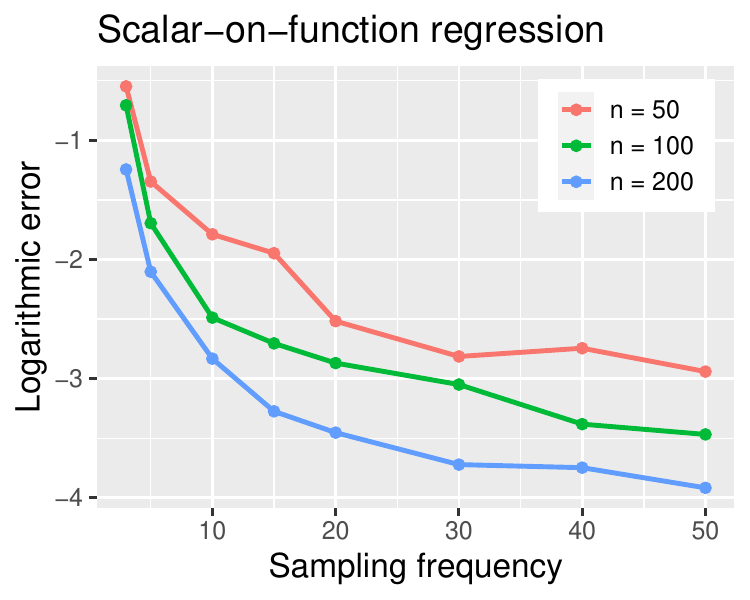}
\end{minipage}
\hfill
\begin{minipage}{0.49\linewidth}
\centering
\includegraphics[width=\linewidth]{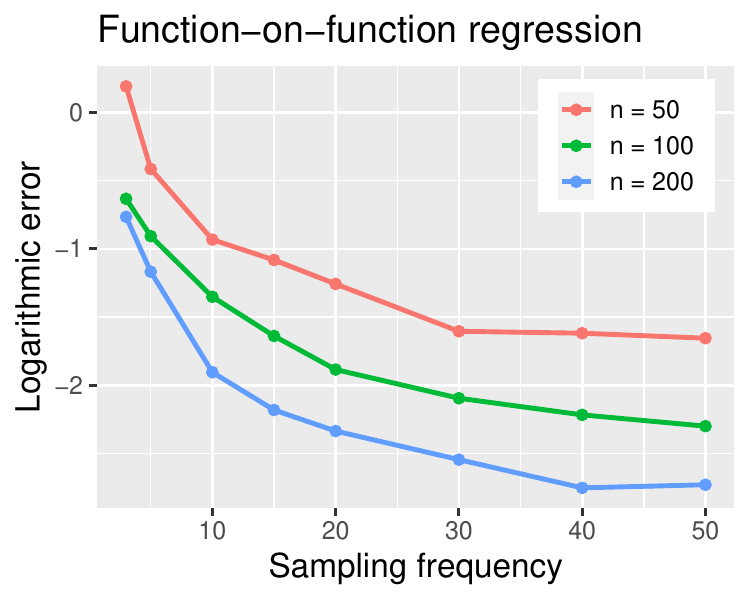}
\end{minipage}
\caption{Logarithmic error under different sampling frequencies. 
The left panel shows the scalar-on-function regression setting and the right panel shows the function-on-function regression in cosine design setting. 
The horizontal axis is the sampling frequency, taking values \(3,5,10,15,20,30,40,50\); for function-on-function regression, we set \(m_x=m_y\). Each line represents a different sample size.}\label{fig:pt_error}
\end{figure}

For each basis design and each combination of \(n\) and \(m_x\), we conduct 100 Monte Carlo simulations.
The average errors for the cosine design are presented in Table \ref{tab:b2v2}, with standard deviations provided in parentheses.
Due to space constraints, the Legendre-design results are reported in the Supplementary Material \citep{supplyment} and lead to qualitatively similar conclusions.
In each row, the method achieving the lowest average error is highlighted in bold. These results demonstrate the strengths of the proposed estimator in all cases compared to the benchmark approaches especially when the sampling frequency is small. 
As the sampling frequency grows, the difference between the proposed method and the RKHS method for fully observed data becomes smaller, consistent with theoretical expectations. Moreover, the standard deviations of the proposed method remain relatively small across all settings, indicating its stability.
These findings emphasize the merits of the proposed method for regression tasks with scalar response involving discretely observed functional data.
A possible reason for the relatively high variability of PACE in sparse settings is that its conditional score calculation involves the inversion of subject-specific covariance matrices. As \(n\) and \(m_x\) increase, such as \(n=100,m_x=6\), the results become more stable and improved.
Finally, the left panel of Figure \ref{fig:pt_error} shows the logarithmic error curve under different \(n\) and \(m_x\). It shows that increasing the sampling frequency substantially improves performance when it is sparse, whereas the improvement becomes limited once the sampling frequency is sufficiently large. This pattern is consistent with the theoretical phase-transition phenomenon: after the sampling frequency passes some transition boundary, further increases yield diminishing benefits. From a practical perspective, this suggests that one may not need to pursue very dense sampling when the cost of increasing the sampling frequency is high. \label{simusf}

\begin{table}[ht]
\centering
\caption{  The Monte Carlo averages and standard deviation (in parentheses) of the error \(\mathcal{E}_{\bo}\) of the proposed method and the other five benchmark methods, for scalar-on-function regression with cosine design. }\label{tab:b2v2}
\resizebox{\textwidth}{!}{
\begin{tabular}{cccccccc}
  \hline
\(n\) & \(m\) & Proposed & FullyRKHS & Plug-in & IN & PACE & MC \\ 
  \hline
\(50\) & \(3\) & \textbf{0.5788} (0.4117) & 1.2675 (0.6028) & 0.9492 (1.1531) & 1.7499 (1.5332) & 1.3767 (2.4667) & 1.4872 (0.8066) \\ 
  \(50\) & \(6\) & \textbf{0.2392} (0.1751) & 0.5440 (0.2946) & 0.5170 (0.4968) & 0.6847 (0.4745) & 0.5239 (1.4602) & 0.6361 (0.3854) \\ 
  \(50\) & \(10\) & \textbf{0.1672} (0.1483) & 0.2742 (0.1741) & 0.2772 (0.2098) & 0.4099 (0.2765) & 0.2683 (0.4861) & 0.2771 (0.1573) \\ 
  \(50\) & \(20\) & \textbf{0.0807} (0.0714) & 0.1244 (0.0805) & 0.1837 (0.1465) & 0.1750 (0.1211) & 0.1087 (0.0931) & 0.1346 (0.0887) \\ 
  \(100\) & \(3\) & \textbf{0.4936} (0.2782) & 1.2094 (0.3722) & 0.5715 (0.6122) & 1.6057 (1.1057) & 1.0182 (2.6082) & 1.2742 (0.4419) \\ 
  \(100\) & \(6\) & \textbf{0.1640} (0.1372) & 0.4371 (0.2129) & 0.3271 (0.2389) & 0.5920 (0.3519) & 0.2528 (0.1944) & 0.4480 (0.1952) \\ 
  \(100\) & \(10\) & \textbf{0.0831} (0.0716) & 0.2325 (0.1097) & 0.1753 (0.1230) & 0.3235 (0.1677) & 0.1356 (0.1376) & 0.2275 (0.1058) \\ 
  \(100\) & \(20\) & \textbf{0.0567} (0.0423) & 0.1080 (0.0615) & 0.0940 (0.0710) & 0.1382 (0.0792) & 0.0693 (0.0713) & 0.1000 (0.0599) \\ 
  \(200\) & \(3\) & \textbf{0.2882} (0.1678) & 1.0559 (0.2491) & 0.4015 (0.2733) & 1.5984 (0.9492) & 0.8313 (2.2677) & 1.0895 (0.2589) \\ 
  \(200\) & \(6\) & \textbf{0.0889} (0.0642) & 0.4322 (0.1249) & 0.2547 (0.1364) & 0.5805 (0.2643) & 0.1963 (0.1348) & 0.4259 (0.1362) \\ 
  \(200\) & \(10\) & \textbf{0.0589} (0.0493) & 0.2030 (0.0672) & 0.1398 (0.1234) & 0.2932 (0.1305) & 0.1024 (0.0901) & 0.1904 (0.0744) \\ 
  \(200\) & \(20\) & \textbf{0.0316} (0.0239) & 0.0866 (0.0358) & 0.0628 (0.0591) & 0.1186 (0.0683) & 0.0400 (0.0428) & 0.0777 (0.0324) \\ 
   \hline
\end{tabular}
}
\end{table}

\subsection{Function-on-function regression}\label{sec:simulationff}
For the function-on-function regression problem, we set the underlying slope function \(\bo\) as 
\[
\bo(s,t) = \sum_{k=1}^{\infty}  4(-1)^{k+1} k^{-2} \phi_{k}(s)\phi_{k}(t).
\]
Then the response function \(Y_i\)'s are generated according to 
\[
Y_{i}(s) = \int_{\mathcal{T}_{x}} X_{i}(t) \bo(s,t) dt + \sum_{k=1}^{\infty} k^{-3} \zeta_{ik} \phi_k(s), \quad 1\leq i \leq n,
\]
where \(\zeta_{ik}\) are independently drawn from \(N(0,1)\). 
Furthermore, the discrete observations are obtained as 
\(
Y_{il} = Y_{i}(S_{il}) + e_{il},\) for \(1\leq i \leq n, 1\leq l \leq m_{y}
\)
with measurement noises \(e_{il}\)'s i.i.d. following \(N(0,0.5^2)\) and \(S_{il}\)'s
independently drawn from the uniform distribution over \(\mathcal{T}_{y}\).  
To implement the proposed method, we set both \(K^x\) and \(K^y\) as the Matérn \(3/2\) kernel, i.e.  
\(
K^x(u_1, u_2) = K^y(u_1, u_2) = (1 + |u_1 - u_2|) \exp(-|u_1 - u_2|),
\)
which further implies the tensor kernel function \(K(s_1,t_1;s_2,t_2)=K^{y}(s_1,s_2)K^{x}(t_1,t_2)\). 
In this part, we consider sample sizes of \( n = 50, 100, 200 \) and examine different sampling scheme with \( m_x =m_y = 3, 6, 10, 20 \) for simplicity. 
To facilitate the computation, the solution is restricted to a subspace of at most \(2500\) dimensions. When \(nm_xm_y\) exceeds this threshold, a random choice of basis functions would be made. Additionally, we generate an independent set of \(500\) pairs i.i.d. functions to evaluate the corresponding estimation error \(\mathcal{E}_{\bo}(\hat{\beta})\).

To assess the performance of the proposed method,
we compare it against several established approaches, including the approximated least-square methods based on FPCA, RKHS method OPFFR-S proposed by \citet{sun2018optimal} and RKHS based FPCA method (referred RKHSPCA latter) proposed by \citet{mostafaiy2019optimal}.
For the approximated least-square methods, we compare multiple score estimation techniques, including the integral approximation method, the conditional expectation method, and the Monte Carlo method. Similar to the scalar-on-function problem, the baseline methods also adopt a standard cross-validation procedure to select tuning parameters, while the candidate parameters of all methods are taken to be \(\Lambda = \{4^{-9},4^{-8},\ldots,4^{-1}\}\).
Table \ref{tab:ffb2v2} presents the results with cosine design for the proposed method alongside five benchmark alternatives, indicating that the proposed approach outperforms the others. See Supplementary Material \citep{supplyment} for the corresponding results under the Legendre design, which lead to qualitatively similar conclusions and further support the proposed method under different potential misalignment.
In particular, compared to the OPFFR-S method, which first recovers the data before applying a regression method designed for fully observed data using grids, the proposed method, which is based on the pooling unbiased operator estimation, is more accurate and stable. 
Moreover, the results demonstrate that the proposed method also performs better than FPCA-based methods. 
Specifically, RKHSPCA performed second only to our method, primarily because they adopted the pooling strategy proposed by \citet{yao2005b} for Gaussian processes. Finally, logarithmic error curve in the right panel of Figure \ref{fig:pt_error} illustrates the diminishing improvement as the sampling frequencies increase, also supporting the theory.

\begin{table}[ht]
\centering
\caption{  The Monte Carlo averages and standard deviation (in parentheses) of the error \(\mathcal{E}_{\bo}\) of the proposed method and the other five benchmark methods, for function-on-function regression with cosine design. }\label{tab:ffb2v2}
\resizebox{\textwidth}{!}{
\begin{tabular}{ccccccccc}
  \hline
\(n\) & \(m_x\) & \(m_y\) & Proposed & OPFFR-S & IN & PACE & MC & RKHSPCA \\ 
  \hline
50 & 3 & 3 & \textbf{1.2120} (2.6020) & 12.4922 (39.4993) & 2.7684 (1.1817) & 2.5760 (3.1839) & 2.1556 (0.9035) & 1.4128 (1.7811) \\ 
  50 & 6 & 6 & \textbf{0.7593} (2.7422) & 7.5348 (18.5352) & 1.1771 (0.4902) & 2.0945 (4.2141) & 1.0853 (0.4371) & 0.8459 (0.5838) \\ 
  50 & 10 & 10 & \textbf{0.3940} (0.1957) & 1.2879 (1.6185) & 0.7344 (0.2530) & 0.6376 (0.4747) & 0.7243 (0.2362) & 0.6871 (0.4528) \\ 
  50 & 20 & 20 & \textbf{0.2846} (0.1850) & 0.4236 (0.2920) & 0.4337 (0.1934) & 0.2868 (0.1461) & 0.4788 (0.2190) & 0.5096 (0.2135) \\ 
  100 & 3 & 3 & \textbf{0.5314} (0.3044) & 4.4701 (6.4807) & 2.3611 (0.8503) & 1.5384 (2.5360) & 1.6284 (0.5184) & 0.8282 (0.5696) \\ 
  100 & 6 & 6 & \textbf{0.4795} (0.8887) & 2.0291 (2.4097) & 0.9392 (0.3231) & 0.6803 (0.6065) & 0.7580 (0.2132) & 0.4932 (0.2038) \\ 
  100 & 10 & 10 & \textbf{0.2590} (0.1169) & 0.6013 (1.1227) & 0.5294 (0.1659) & 0.3703 (0.1575) & 0.5593 (0.1858) & 0.5015 (0.2303) \\ 
  100 & 20 & 20 & \textbf{0.1519} (0.0763) & 0.2117 (0.1017) & 0.2831 (0.1137) & 0.1542 (0.0709) & 0.3288 (0.1464) & 0.4323 (0.1624) \\ 
  200 & 3 & 3 & \textbf{0.4655} (0.2678) & 7.2745 (15.5985) & 2.1975 (0.6514) & 1.2264 (2.1211) & 1.3468 (0.3186) & 0.5028 (0.2203) \\ 
  200 & 6 & 6 & 0.4488 (1.7489) & 1.6758 (2.6608) & 0.8164 (0.2430) & 0.5426 (0.4201) & 0.6613 (0.1706) & \textbf{0.4335} (0.1674) \\ 
  200 & 10 & 10 & \textbf{0.1490} (0.0793) & 0.6678 (2.9211) & 0.4451 (0.1286) & 0.3198 (0.7406) & 0.4512 (0.1185) & 0.3961 (0.1159) \\ 
  200 & 20 & 20 & \textbf{0.0968} (0.0532) & 0.1788 (0.1697) & 0.2384 (0.0977) & 0.1340 (0.2143) & 0.2505 (0.1074) & 0.3186 (0.0968) \\ 
   \hline
\end{tabular}
}
\end{table}

\section{Real data}\label{sec:realdata}
To demonstrate the advantages of the proposed method, we apply it to two examples, including the wheat data and the CONTENT growth data. 
Specifically, the first dataset is considered in a scalar-on-function regression model.
Since this data is densely observed, we employ a random sparsification procedure to examine the performance of the proposed method at various sampling frequencies.
Moreover, the second dataset is longitudinal and sparsely collected, and we consider the growth prediction in the function-on-function regression model, which is a typical application scenario for functional data.

\subsection{The wheat dataset}
In this subsection, we illustrate the application of the proposed methodology to the wheat dataset introduced by \citet{kalivas1997two}. 
The dataset contains near-infrared (NIR) spectral measurements of wheat samples, along with the corresponding protein content values. It includes data from \(100\) individuals, with NIR spectrum of each subject recorded at \(701\) wavelengths, spaced \(2\) nm apart, ranging from \(1100\) nm to \(2500\) nm. The primary goal of this study is to regress the protein concentration on the NIR spectra, investigating the functional relationship between wheat protein content and spectral patterns.

To evaluate the estimation performance, we randomly split the dataset into training and testing sets, consisting of 80 and 20 subjects, respectively. Note that this dataset is densely observed, we simulate different levels of data sparsity by randomly selecting \(m = 5, 10, 20\) measurements per subject in the training set. To reduce the impact of unnecessary randomness, we repeat the experiments 100 times, re-partitioning the training and testing sets and resparsifying the training data each time.  
The performance of the proposed method, in comparison to five alternative approaches, is summarized in Table \ref{tab:wheat}. 
Across various sampling frequencies \( m_x \), the proposed approach achieves the lowest empirical mean error, with significant improvements over the other methods. This suggests that our approach effectively leverages available functional information, particularly in addressing the challenges posed by sparse measurements.
{Additionally, we note that FullyRKHS perform relatively well in this sparse setting. One possible reason is that the data exhibits smooth spectral patterns, hence the integration approximation using artificially sparsified observations yields good results. This might explain why MC outperforms other scores estimators. 
The weaker performance of PACE relative to IN may be related to the Gaussian working assumptions required by PACE; which may also be less appropriate for these data.}\label{51jieshi}

\begin{table}[ht]
\centering
\caption{  The Monte Carlo averages and standard deviation (in parentheses) of the prediction error of the proposed method and the other five benchmarks, for the scalar-on-function regression on the wheat dataset at different sparsity.}\label{tab:wheat}
\resizebox{\textwidth}{!}{
\begin{tabular}{cccccccc}
  \hline
\(m\) & Proposed & FullyRKHS & Plug-in & PACE & IN & MC \\ 
  \hline
 \(5\) & \textbf{0.8029} (0.3514) & 0.8261 (0.3514) & 0.9024 (0.3777) & 0.9218 (0.3737) & 0.8701 (0.3792) & 0.8231 (0.3435) \\ 
 \(10\) & \textbf{0.6761} (0.2905) & 0.7491 (0.3128) & 0.8793 (0.3909) & 0.8764 (0.3733) & 0.7955 (0.3566) & 0.7190 (0.3048) \\ 
\(20\) & \textbf{0.5920} (0.2562) & 0.6696 (0.2758) & 0.7351 (0.3782) & 0.6715 (0.3165) & 0.7123 (0.3318) & 0.6271 (0.2513) \\ 
   \hline
\end{tabular}
}
\end{table}

\subsection{The CONTENT dataset}

The CONTENT child growth study is a longitudinal cohort study that was originally designed to investigate the impact of Helicobacter pylori infection on child growth.
To identify expectant mothers and infants under three months of age, an initial demographic survey was conducted. Participants were randomly selected from the Las Pampas de San Juan Miraflores and Nuevo Paraíso regions between May 2007 and February 2011. 
Anthropometric data were collected from 197 children \citep{jaganath2014first}, with visit schedules varying across participants.

\begin{figure}[htbp]
    \centering
    \includegraphics[width=1\textwidth]{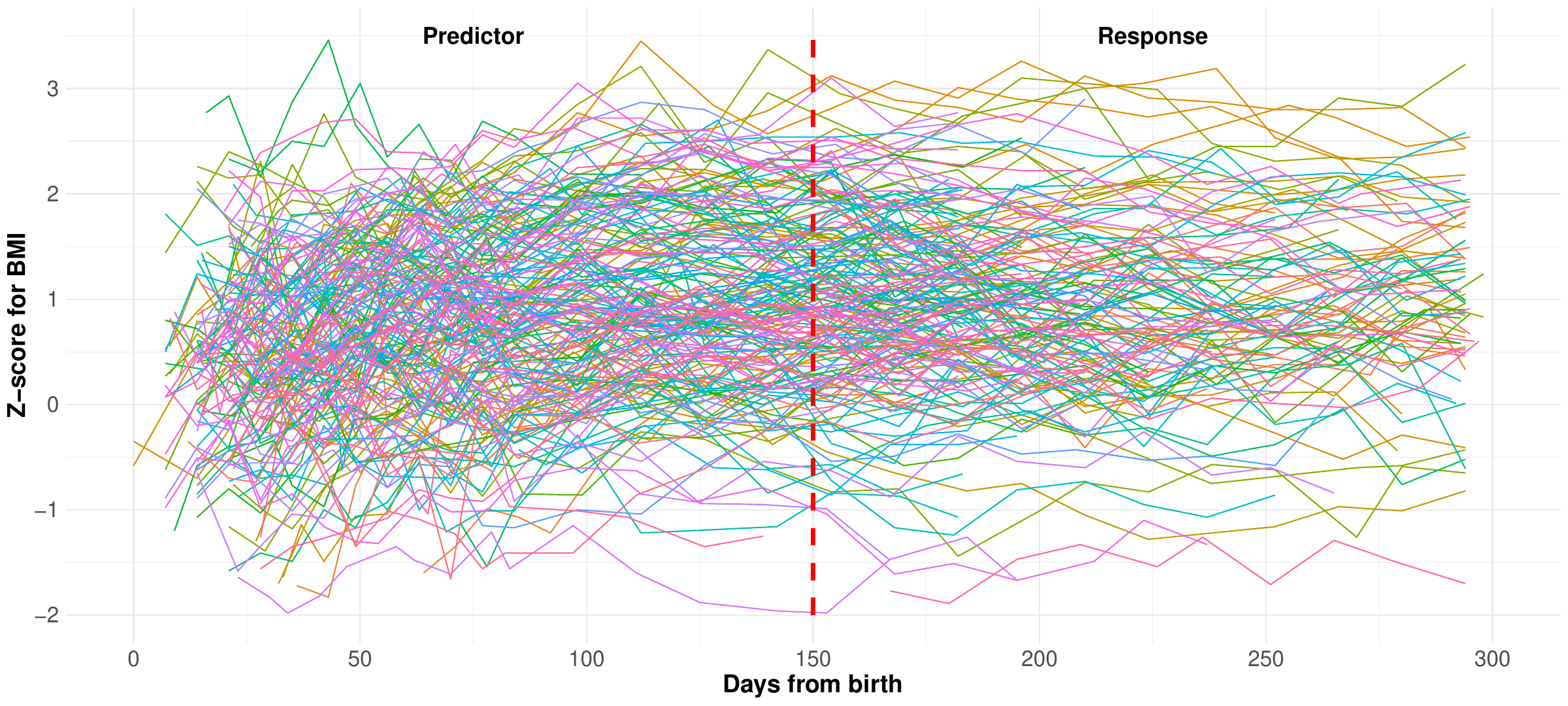}
    \caption{Longitudinal observation of Z-score for BMI as a function of day from birth. 
    Each colored thin line corresponds to an individual. %
    The red vertical dotted line separates the past and future, which serve as predictors and responses in this study. 
    }
    \label{fig:vis_content_data}
\end{figure}

The objective of this study is to utilize the BMI Z-scores collected during the initial 150 days after birth as predictor functions to predict that in the subsequent period, from day 151 to day 300. This setup involves modeling the functional relationship between the growth trajectory in different periods, with the challenge of the irregularity and sparsity of the measurements. 
Among all subjects, the number of observations for the predictor functions ranged from \(5\) to \(17\), with a median of \(14\), while the sampling frequency for the response functions varied from \(2\) to \(11\), with a median of \(7\).  
Figure \ref{fig:vis_content_data} provides a visual representation of the data. %

To evaluate the performance of the proposed methodology, similar to Section \ref{sec:simulationff}, we conduct a comparison with various baseline methods, including the OPFFR-S method, the RKHS-PCA comparison method, and the approximated least-square methods based on three kinds of score estimators.
After centering the data using the estimated mean function, we randomly divide the full data into training and test sets.
Given the sparsity of the data, the test set is taken to contain twice as many individuals as the training set to reduce variance in evaluation. 
Then we set 100 repetition experiments to further reduce the randomness introduced by set partitioning.
The results, summarized in Table \ref{tab:content}, present the mean, standard deviation, and three quartiles of the empirical prediction errors across all repetitions for the proposed method and five alternative approaches. The best-performing method for each criterion is highlighted in bold.
The table shows that the proposed method consistently outperforms the alternatives across these evaluation metrics. 
In comparison, the proposed method, which is based on a pooling estimated operator, is better and more stable than the OPFFR-S method which recovers each curve first and then applies the traditional regression method designed for fully observed data.
Similar conclusion can be drawn when compared with FPCA-based methods using three versions of score estimators. Furthermore, since the Gaussian assumption for the real data no longer holds, the performance of RKHSPCA is inferior to that of the simulation.

\begin{table}[ht]
\centering
\caption{  
The mean, standard deviation, and three quartiles of the empirical error of the proposed method and five alternative approaches, for 100 repeated experiments on the training/testing sets partition. The best result of each column is in boldface.
}\label{tab:content}
\centering
\begin{tabular}{cccccc}
  \hline
Method & Mean & Median & Sd & firstQu & thirdQu \\ 
  \hline
Proposed & \textbf{0.3536} & \textbf{0.2754} & \textbf{0.1972} & \textbf{0.1868} & \textbf{0.4899} \\ 
  OPFFR-S & 0.4992 & 0.3782 & 0.3430 & 0.2458 & 0.6890 \\ 
  IN & 0.4271 & 0.3517 & 0.2362 & 0.2324 & 0.5802 \\ 
  PACE & 0.4024 & 0.3072 & 0.2258 & 0.2093 & 0.5926 \\ 
  MC & 0.3823 & 0.3036 & 0.2171 & 0.1916 & 0.6020 \\ 
  RKHSPCA & 0.5258 & 0.4113 & 0.2842 & 0.2834 & 0.7647 \\ 
   \hline
\end{tabular}
\end{table}

\begin{supplement}
\stitle{Supplement to ``Functional linear regression from sparse to dense designs: a pooling-ridge method and minimax optimality''}
\sdescription{Contains the proofs of the main theoretical results and technical lemmas, together with additional simulation results.}
\end{supplement}

\bibliographystyle{imsart-nameyear} %
\bibliography{bibliography}       %

\end{document}